\documentclass[aps,prd,10pt,english,reprint,showpacs]{revtex4-1}

\usepackage{babel}
\usepackage{mathptmx}
\usepackage{amssymb,textcomp}
\usepackage{xcolor}
\usepackage{amsmath,graphicx,array}
\usepackage{tikz}
\usetikzlibrary{patterns}

\definecolor{blau1}{rgb}{0,0.12,0.50}   
\definecolor{blau2}{rgb}{0.6,0.69,1}    
\definecolor{blau3}{rgb}{0.87,0.90,1}   
\definecolor{blau4}{rgb}{0,0.062,0.27}  
\definecolor{orange}{rgb}{1,0.88,0.50}  
\definecolor{rot}{rgb}{0.79,0.31,0}     

\newcolumntype{C}{>{$}c<{$}}
\newcommand{\M}[1]{\multicolumn{1}{m{13mm}}{\centering #1}}

\newcommand{\circc}[1]{%
\begin{tikzpicture}\draw[fill=#1, color=#1] (0,0) circle (.1);\end{tikzpicture}}

\newcommand{\circpdash}[1]{%
\begin{tikzpicture}\draw[color=#1,dashed,dash pattern=on 1.5pt off 1.5pt] (0,0) circle (.1);\end{tikzpicture}}

\newcommand{\rectc}[1]{%
\begin{tikzpicture}\draw[fill=#1, color=#1]	(0,0) rectangle (.4,.2);\end{tikzpicture}}

\newcommand{\rectr}[1]{%
\begin{tikzpicture}%
\draw[color=#1, pattern=north east lines, pattern color=#1]	(0,0) rectangle (.4,.2);%
\end{tikzpicture}}
\newcommand{\rectl}[1]{%
\begin{tikzpicture}%
\draw[color=#1, pattern=north west lines, pattern color=#1] (0,0) rectangle (.4,.2);%
\end{tikzpicture}}

\newcommand{\R}{\mathbb{R}}
\newcommand{\Lag}{\mathcal{L}}
\newcommand{\K}{\mathcal{K}}

\let\blqq=\textquotedblleft
\let\brqq=\textquotedblright

\makeatletter
\newcommand{\drm}{\@ifstar{\mathrm{d}}{\:\mathrm{d}}}
\makeatother

\graphicspath{{Bilder_pdf/}}

\begin{document}

\title{Photon Regions and Shadows of Kerr--Newman--NUT Black Holes
with a Cosmological Constant}
\author{Arne Grenzebach}
\author{Volker Perlick}
\author{Claus L\"ammerzahl}
\affiliation{ZARM, Universit\"at Bremen, Am Fallturm, D-28359 Bremen, Germany}
\date{\today} 

\begin{abstract}
We consider the Pleba{\'n}ski class of electrovacuum solutions to 
the Einstein equations with a cosmological constant. These space-times, 
which are also known as the Kerr--Newman--NUT--(anti-)de Sitter space-times,
are characterized by a mass $m$, a spin $a$, a parameter $\beta$ that
comprises electric and magnetic charge, a NUT parameter $\ell$ and a 
cosmological constant $\Lambda$. Based on a detailed discussion of the photon 
regions in these space-times (i.e., of the regions in which spherical 
lightlike geodesics exist), we derive an analytical formula for the 
shadow of a Kerr--Newman--NUT--(anti-)de Sitter black hole, for an 
observer at given Boyer--Lindquist coordinates $(r_O, \vartheta_O)$ in
the domain of outer communication. We visualize the photon regions and 
the shadows for various values of the parameters.
\end{abstract}

%
%
%
%
%
%

\maketitle

\section{Introduction}\label{sec:intro}
Over the last twenty years observations have produced increasing
evidence for the existence of a supermassive black hole at the center
of our galaxy. This evidence comes from the observation of orbits of
stars in the infrared \cite{EckartGenzel.1996,GillessenEisenhauerEtAl.2009}
which allows to estimate the mass of the central object. In
combination with estimates of the volume in which this mass must be
concentrated the result strongly supports the hypothesis of a black
hole. These observations are expected to become even more precise when
the GRAVITY instrument \cite{EisenhauerEtAl.2009} goes into operation
soon. In addition, it is planned to explore the inner region of the
center of our galaxy, in the order of magnitude of the Schwarzschild
radius of the central mass, with submillimeter radio telescopes. From
this project, which is called the Event Horizon Telescope
\cite{DoelemanWeintroub.2008}, we expect a radio image of the shadow
of the central black hole in a few years' time. Therefore, it is
timely to advance the theoretical investigations of the shadows of
black holes as far as possible, as a basis for evaluating the
observational results that are to be expected soon.

\begin{figure}[bp]
	\centering
	\includegraphics{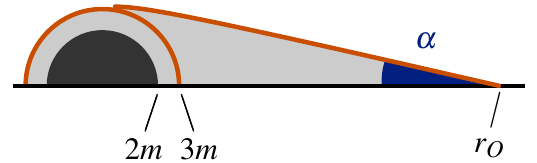}
	\caption{Angular radius $\alpha$ of the shadow of a Schwarzschild 
	black hole, given by Synge's formula, Eq.~\eqref{eq:Synge}.}
	\label{fig:Synge}
\end{figure}

For an observer at radius coordinate $r_O$ in the 
Schwarzschild space-time, the shadow can be constructed in the 
following way. We assume that there are light sources 
distributed on the sphere $r=r_L$ for some chosen $r_L>r_O$. We 
consider all light rays issuing from the observer's
position into the past. Some of them will reach a light 
source at $r_L$, after being deflected by the black hole; to 
the initial directions of this first class of light rays we 
associate brightness on the observer's sky. Some of them will 
go to the horizon and never reach a light source at $r_L$; to the
initial directions of this second class of light rays we 
associate darkness on the observer's sky. The second class fills
the shaded region in Fig.~\ref{fig:Synge}. The borderline between 
the two classes are light rays that asymptotically spiral towards 
the photon sphere at $r=3m$ (with $G=1$, $c=1$). 
Therefore, in this case the shadow is circular and its angular radius
is determined by light rays that approach the photon sphere, see
again Fig.~\ref{fig:Synge}. 
For simplicity, we have constructed the shadow with 
light sources on a sphere $r=r_L$. From the geometry it is clear
that we could have light sources anywhere else as long as they
are outside of the shaded region in Fig.~\ref{fig:Synge}.

Synge \cite{Synge.1966} was the first to
calculate what we nowadays call the shadow of a Schwarzschild black
hole.  (Synge did not use the word \blqq shadow\brqq\ but he
investigated the condition under which photons could escape to
infinity.)  He found that the angular radius $\alpha$ of the shadow is
given by the simple formula
\begin{equation}
	\sin^2 \alpha = \frac{27}{4} \; \frac{(\rho_O -1)}{\rho_O^{3}}
\label{eq:Synge}
\end{equation}
where $\rho_O = r_O/(2m)$ is the ratio of the observer's $r$ coordinate 
$r_{O}$ and the Schwarzschild radius. For the
black hole at the galactic center, an observer on the Earth is at
$r_O \approx 8.3$ kpc, and the mass is $m \approx 4.1 \times 10^{6}$
Solar masses \cite{GhezSalim.2008,GillessenEisenhauerEtAl.2009}. If
one inserts these values into Synge's formula one gets an angular
radius of $\alpha \approx 25$ microarcseconds which is expected to be
resolvable with Very Long Baseline Interferometry (VLBI) soon
\cite{DoelemanWeintroub.2008,HuangCai.2007}.

For a Kerr black hole, there is no longer a photon sphere and the
shadow is no longer circular. The photon sphere breaks into a \blqq
photon region\brqq\ which is filled by spherical lightlike geodesics,
i.e. by lightlike geodesics each of which is confined to a sphere
$r=\mathrm{constant}$. The boundary of the shadow corresponds to light
rays that asymptotically spiral towards one of these spherical
lightlike geodesics. The deviation of the shadow from a circle is a
measure for the spin of the black hole. Bardeen \cite{Bardeen.1973}
was the first to correctly calculate the shadow of a Kerr black hole,
the results can also be found, e.g., in Chandrasekhar's book
\cite{Chandrasekhar.1983}. For pictures of individual spherical
lightlike geodesics in the Kerr space-time we refer to Teo
\cite{Teo.2003}, and for a discussion and a picture of the photon
region in the Kerr space-time to Perlick \cite{Perlick.2004}.

The shadow has also been discussed for other black holes (and for
naked singularities), e.g. for the Kerr--Newman space-time
\cite{Vries.2000}, for $\delta=2$ Tomimatsu-Sato space-times
\cite{BambiYoshida.2010}, for black holes in extended Chern--Simons
modified gravity \cite{AmarillaEiroa.2010}, in a Randall--Sundrum
braneworld scenario \cite{AmarillaEiroa.2012}, and a Kaluza--Klein
rotating dilaton black hole \cite{AmarillaEiroa.2013}, 
for the Kerr--NUT space-time \cite{AbdujabbarovAtamurotov.2012}, 
for multi-black holes \cite{YumotoNitta.2012}, and for regular 
black holes \cite{LiBambi.2014}. Hioki and Maeda \cite{HiokiMaeda.2009}
introduced a deformation parameter that characterizes the deviation of
the shadow from a circle. Special interest has been devoted to the
question of whether the shadow of a black hole can be used as a test
of the no-hair theorem, see Johannsen and Psaltis
\cite{JohannsenPsaltis.2011}. All these articles are largely based
on ray tracing in the respective space-times, rather than on analytical
studies of the geodesic equation, and they assume that the observer is
at infinity.

In this paper we want to extend the discussion of the shadow in
various directions. First, we consider a class of space-times for which
the shadow has not yet been calculated, namely the Pleba{\'n}ski class
\cite{Plebanski.1975}. The metrics in this class, which are also known
as the Kerr--Newman--NUT--(anti-)de Sitter metrics, depend on five
parameters: A mass $m$, a spin $a$, a parameter $\beta$ that comprises
an electric and a magnetic charge, a NUT parameter $\ell$, and a
cosmological constant $\Lambda$. It is a subclass of the
Pleba{\'n}ski--Demia{\'n}ski class \cite{PlebanskiDemianski.1976} of
stationary axisymmetric type D electrovacuum solutions of Einstein's
field equations with a cosmological constant; the latter includes, in
addition to the five parameters of the Pleba{\'n}ski class, also a
so-called acceleration parameter; in the present work we 
will not consider the acceleration parameter but we are planning 
to study its influence in a separate publication.
Second, we develop the formalism for an observer
\emph{not} at infinity but rather at some given Boyer--Lindquist
coordinates $(r_O,\vartheta_O)$ in the domain of outer communication.
This is essential for the case $\Lambda \neq 0$ because then the
space-time is no longer asymptotically flat and in the case $\Lambda
>0$ the domain of outer communication is separated from $r= \infty$ by
a cosmological horizon. Third, our treatment is fully analytical
rather than based on ray tracing. In particular, we give an exact
analytical formula for the boundary curve of the shadow. We feel that
this is a major advantage because it can serve as a basis for
calculating parameters of the space-time from the shape of the shadow
by analytical means. Fourth, our investigation includes a detailed
discussion of the photon regions in the space-times under
consideration. This is a crucial prerequisite for deriving the
analytical formula of the shadow, and it is also of some interest in
itself.

We emphasize that, as in all the theoretical papers cited above, our
calculation of the shadow is based on the assumptions that light rays
are lightlike geodesics and that there are no light sources near the 
black hole. In view of the black hole at the center of our galaxy
these assumptions are highly idealized. Light rays near the central
black hole are expected to be affected by scattering, and there is
good evidence for the existence of a luminous accretion disk around
the black hole. The effect of scattering on the visibility of the
shadow was numerically demonstrated by Falcke, Melia and Agol
\cite{FalckeMelia.2000}. The visual appearance of an accretion disk
was studied with the help of various ray-tracing programs by several
authors, following the pioneering work of Bardeen and Cunningham
\cite{BardeenCunningham.1973} and Luminet \cite{Luminet.1979}, see
e.g. Dexter et.al. \cite{DexterAgol.2012} or 
Mo{\'s}cibrodzka et.al. \cite{MoscibrodzkaShiokawa.2012}.
A broad overview of observations as well as simulations of phenomena
for the black hole in the center of our galaxy near Sgr~A* is given 
by Dexter and Fragile in \cite{DexterFragile.2013}. %
Whereas the effects of matter certainly have to be taken into
account for a realistic prediction of what will be observed,
calculating the geometrical shadow is of major importance because 
it serves as the basis for all later refinements.

The paper is organized as follows. In Section~\ref{sec:KNUT} we
summarize the relevant properties of space-times of 
the Pleba{\'n}ski class.  In Section~\ref{sec:pregions}
we determine the photon regions for black-hole space-times
of this class. In Section~\ref{sec:shadow} we derive an analytical 
formula, in parameter form, for the boundary curve of the shadow
of such a black hole, as it is seen by an observer with a specified 
four-velocity $e_0$ somewhere in the domain of outer communication.
The results of Sections~\ref{sec:pregions} and \ref{sec:shadow}
are illustrated with several pictures.

\section{The Kerr--Newman--NUT--(anti-)de Sitter metric}\label{sec:KNUT}

The Kerr--Newman--NUT--(anti-)de Sitter space-times are stationary,
axially symmetric type D solutions of the Einstein--Maxwell equations
with a cosmological constant. This class of space-times was introduced
by Pleba{\'n}ski \cite{Plebanski.1975} in 1975. A slightly larger
class, which includes in addition the so-called acceleration
parameter, was found by Pleba{\'n}ski and Demia{\'n}ski
\cite{PlebanskiDemianski.1976} in 1976. For the case without a
cosmological constant, these metrics can be traced back to Carter
\cite{Carter.1968b} and, in the Boyer--Lindquist coordinates we will
use in the following, to Miller \cite{Miller.1973}. A fairly detailed
discussion of the Pleba{\'n}ski\mbox{(--De}mia{\'n}ski) metrics can be found
in the book by Griffiths and Podolsk{\'y}
\cite{GriffithsPodolsky.2009}, see also Stephani et al. 
\cite{StephaniKramer.2003}.

In Boyer--Lindquist coordinates $(t, r, \vartheta , \varphi)$ the 
Pleba{\'n}ski metric is given by \citep[p. 314]{GriffithsPodolsky.2009}
\begin{equation}
\begin{aligned}
	g_{\mu \nu} \drm x^{\mu} \drm x^{\nu} &= 
		\Sigma \bigl( \tfrac{1}{\Delta_{r}}\drm r^2 
		+ \tfrac{1}{\Delta_{\vartheta}}\drm \vartheta^2 \bigr) \\
	&\quad + \tfrac{1}{\Sigma} 
		\bigl( (\Sigma + a\chi)^2 \Delta_{\vartheta}\sin^2\vartheta 
			- \Delta_{r} \chi^2 \bigr) \drm \varphi^2 \\
	&\quad + \tfrac{2}{\Sigma} 
		\bigl( \Delta_{r}\chi - a(\Sigma + a\chi) 
			\Delta_{\vartheta}\sin^2\vartheta \bigr) \drm t \drm \varphi \\
	&\quad - \tfrac{1}{\Sigma} 
		\bigl( \Delta_{r} - a^2\Delta_{\vartheta}\sin^2\vartheta \bigr) \drm t^2
\end{aligned}
	\label{eq:Metric}
\end{equation}
where we use the abbreviations 
\begin{equation}
\begin{aligned}
	\Sigma &= r^2 + \bigl(\ell + a\cos\vartheta \bigr)^2, \\
	\chi   &= a\sin^2\vartheta - 2\ell(\cos\vartheta + C), \\
	\Delta &= r^2 - 2mr + a^2 -\ell^2 + \beta, \\
	\Delta_{r} &= \Delta - \Lambda\bigl( (a^{2}-\ell^{2})\ell^{2} + 
			(\tfrac{1}{3}a^{2}+2\ell^{2})r^{2} +\tfrac{1}{3}r^{4} \bigr),  \\
	\Delta_{\vartheta} &= 1 + \Lambda\bigr( \tfrac{4}{3}a\ell\cos\vartheta 
			+ \tfrac{1}{3}a^{2}\cos^{2}\vartheta \bigl).
\end{aligned}
	\label{eq:MetricFunc}
\end{equation}
Here, rescaled units are used so that the speed of light and 
the gravitational constant are normalized ($c=1$, $G=1$). %
The coordinates $t$ and $r$ range over $]{-}\infty, \infty[\,$, 
while $\vartheta$ and $\varphi$ are standard coordinates on
the two-sphere.
The metric depends on five parameters, namely the mass $m$, the spin
$a$, a parameter $\beta$ for electric and magnetic charge
($\beta=q_{e}^{2}+q_{m}^{2}$), the NUT parameter $\ell$ which is to be
interpreted as a gravitomagnetic charge, and the cosmological constant
$\Lambda$. In addition, there is a parameter $C$ that was introduced
by Manko and Ruiz \citep{MankoRuiz.2005} for modifying the singularity
that is produced by $\ell$ on the $z$ axis, see below. In principle,
the parameters $m$, $a$, $\ell$, $\beta$, $\Lambda$ and $C$ can take
all values in $\R$, although not all combinations are
physically meaningful. Note that for $\beta <0$ the metric cannot be
interpreted as a solution to the Einstein--Maxwell equations, because
in this case the electric or magnetic charge has to be imaginary.
Nonetheless, the case $\beta <0$ is of interest because metrics of
this form occur in some braneworld models, see
\cite{AlievGumrukcuoglu.2005}.

The Pleba{\'n}ski class of metrics contains the Schwarzschild
($a=\beta = \ell=\Lambda=0$), Kerr ($\beta=\ell=\Lambda=0$),
Reissner--Nordstr{\"o}m ($a=\ell=\Lambda=0$), Kottler or
Schwarzschild--\mbox{(anti-)}de Sitter ($a=\beta=\ell=0$), Kerr--Newman
($\ell=\Lambda=0$), and Taub--NUT ($a=\beta=\Lambda=0$) metrics as
special cases.

The metric \eqref{eq:Metric} becomes singular if $\Sigma =0$,
$\Delta_r =0$, $\Delta_{\vartheta} =0$ or $\sin \vartheta =0$.
Some of these singularities are mere coordinate singularities, but some of 
them are true (curvature) singularities. As this issue is of some relevance
for our purpose, we briefly discuss the four types of singularities in the
following paragraphs.

\begin{itemize}
	\item[(a)] $\Sigma =0$. 
	The equation $\Sigma = 0$  is equivalent to 
	\begin{equation}
		r=0 \quad \text{and} \quad \cos\vartheta = -\ell/a.
	\label{eq:RingSing}
	\end{equation}
	If $\ell ^2 < a^2$, this condition is satisfied on a ring. The singularity 
	on this ring turns out to be a true (curvature) singularity if $m \neq 0$.
	One usually refers to it as to the \emph{ring singularity}. Note that, apart 
	from the ring singularity, the sphere $r=0$ is regular. Observers 
	can move through either of the two hemispheres (\blqq throats\brqq)
	that are bounded by the ring singularity, thereby travelling from
	the region $r>0$ to the region $r<0$ or vice versa. 

	If $\ell^2 > a^2$, there is no ring singularity. $\Sigma$ is everywhere
	different from zero and the entire sphere $r = 0$ is regular. 

	In the borderline case $\ell^2 = a^2$ the ring singularity degenerates
	into a point on the axis. The case $\ell = a = 0$ is special because
	in this case the entire sphere $r=0$ degenerates into a point singularity 
	that separates the region $r>0$ from the region $r<0$. In this case we 
	have two disconnected space-times. 

	\item[(b)] $\Delta_r =0$. 
	If we exclude the case $a= \ell =0$, each zero of $\Delta_{r}$ on 
	the real line, $- \infty < r < \infty$, is a coordinate singularity	
	which indicates a horizon. As $\Delta_{r}$ is a fourth-order 
	polynomial of $r$ with real coefficients, the number 
	of horizons can be 4, 2 or 0, where zeros of $\Delta_r$ have 
	to be counted with multiplicity. We say that the horizon at
	the biggest $r$ coordinate is the first horizon, the next 
	one is the second, and so on. 

	If $\Lambda \leq 0$, the second derivative of $\Delta_r$ with
	respect to $r$ is strictly positive. Therefore, the number of
	zeros of $\Delta_r$ is either 2 or 0. In the first case we
	have a black hole, in the second case a naked singularity or a 
	regular space-time. In the black-hole case, the region between 
	$r= \infty$ and the first horizon is called the \emph{domain of 
	outer communication} of the black hole. This is the region where 
	we will place our observers for observing the shadow of the black 
	hole. On the domain of outer communication, the vector field 
	$\partial_r$ is spacelike which is equivalent to $\Delta_{r}>0$.
	If $\Lambda = 0$, the equation  $\Delta_{r}=0$ reduces from
	fourth to second order. In this case the horizons are at 
	\begin{equation}
		r_{\pm} = m \pm \sqrt{m^2 -a^2 +\ell^2 - \beta}
		\label{eq:InOutHorizon}
	\end{equation}
	if $a^{2} \le a_{\mathrm{max}}^{2} := m^{2}+\ell^{2}-\beta$; 
	if $a^{2} > a_{\mathrm{max}}^{2}$ there are no horizons, i.e., we
	have a naked singularity or a regular space-time. 

	If $\Lambda >0$, the vector field $\partial_r$ is timelike for 
	big values of $r$. Therefore, the first horizon, if it exists, is
	a cosmological horizon. We have a black hole if there are four
	horizons altogether. The \emph{domain of outer communication} is the 
	region between the first and the second horizon. Again, the vector 
	field $\partial_r$ is spacelike on the domain of outer communication. 
	As in the case $\Lambda \leq 0$, we will restrict ourselves to the 
	black-hole case and we will place our observers in the 
	domain of outer communication.

	\item[(c)] $\Delta_{\vartheta}=0$. 
	If $\Lambda \neq 0$, it is possible that zeros of 
	$\Delta_{\vartheta}$ occur at values $\cos^2 \vartheta < 1$. 
	In close analogy to the zeros of $\Delta_r$, 
	any such zero of $\Delta _{\vartheta}$ is a coordinate 
	singularity which indicates a horizon. In 
	this case, the horizon is situated on a cone $\vartheta 
	= \mathrm{constant}$ rather than on a sphere $r = \mathrm{constant}$.
	The vector field $\partial_{\vartheta}$ changes its causal
	character from spacelike to timelike when such a horizon is 
	crossed.  This situation is hardly of any physical relevance. 
	Therefore, we want to choose the parameters such that it is
	excluded. A sufficient condition can be found in the following
	way. The equation $\Delta_{\vartheta} = 0$
	leads to a quadratic equation for $a \cos \vartheta$ with 
	solution
	\begin{equation}
		a \cos\vartheta_{\pm} = -2\ell \pm \sqrt{4\ell^{2}-3/\Lambda}.
		\label{eq:ConicSing}
	\end{equation}
	Therefore, if we restrict ouselves to values of $\ell$ and 
	$\Lambda$ such that 
	\begin{equation}\label{eq:ellLambda}
		4 \ell^2 \Lambda < 3
	\end{equation}
	we can be sure that $\Delta_{\vartheta}$ has no zeros.

	\item[(d)] $\sin \vartheta =0$. 
	The metric has a singularity on the axis $\sin \vartheta =0$, as is
	always the case when using spherical polar coordinates. If $\ell
	\neq 0$, however, this is not just a coordinate singularity but
	rather a true singularity. By choosing the Manko--Ruiz parameter 
	$C$ appropriately one can decide on which part of the axis the 
	singularity is situated.

	To demonstrate this, we observe that in the limit 
	$\cos\vartheta\to \pm 1$ we have $\Sigma \to r^2+
	(\ell \pm a)^2$ and $\chi \to -2\ell(\pm1 + C)$. As
	a consequence, the metric coefficient
	\begin{equation}
		g^{tt} = 
		\frac{\chi^{2}}{\Sigma \Delta_{\vartheta} \sin^{2}\vartheta}
		-\frac{(\Sigma + a\chi)^{2}}{\Sigma \Delta_{r}}
		\label{eq:AxialSing}
	\end{equation}
	diverges unless $C = \mp 1$. This divergent behavior indicates that 
	either the coordinate function $t$ or the metric $g$ becomes pathological.
	It was shown by Misner \cite{Misner.1963} that this singularity can be 
	removed if one makes the time coordinate $t$ periodic. (Misner restricted
	himself to the Taub--NUT metric, $a=\beta=\Lambda=0$, with $C=1$ but his
	reasoning applies equally well to the general case.) We do \emph{not} 
	follow this suggestion because it leads to a space-time with closed 
	timelike curves through \emph{every} event. Instead, we adopt Bonnor's 
	interpretation \cite[p.~145]{Bonnor.1969} of the axial singularity who 
	viewed it as a \blqq massless source of angular momentum\brqq. For $C=1$, 
	the singularity is on the half-axis $\vartheta =0$, for $C=-1$ it is on the 
	half-axis $\vartheta = \pi$ and for any other value of $C$ it is on both
	half-axes. Note that each half-axis extends from $r= -\infty$ to 
	$r = \infty$.

	Metrics \eqref{eq:Metric} with different values of $C$ are
	\emph{locally} isometric near all points off the axis. This follows
	from the fact that a coordinate transformation $t'=t-2 \ell 
	\widetilde{C}\varphi$ yields, again, a metric \eqref{eq:Metric} with 
	$C'=C+\widetilde{C}$. With the help of such a coordinate transfomation 
	with $\widetilde{C}=-C$, the parameter $C$ can be eliminated from the 
	geodesic equation, see Kagramanova et al. \cite{KagramanovaEtAl.2010}.
	Note, however, that this transformation does not work globally because 
	$\varphi$ is periodic and $t$ is not, and it does not work near the 
	axis because $\varphi$ is pathological there.

	Moreover, a coordinate transformation $\big( t',r' , \vartheta' , \varphi' \big) 
	= \big( t , r , \pi - \vartheta ,  -\varphi \big)$ transforms a metric 
	\eqref{eq:Metric} into a metric of the same form, but with the signs
	of $\ell$ and $C$ inverted. This demonstrates that a metric with 
	parameters $(m,a,\Lambda,\beta,\ell,C)$ is \emph{globally} isometric
	to a metric with parameters $(m,a,\Lambda,\beta,-\ell,-C)$.
\end{itemize}

We have seen that the vector fields $\partial_r$ and
$\partial_{\vartheta}$ change their causal character from spacelike to
timelike if a horizon is crossed. The vector fields $\partial_t$ and
$\partial_{\varphi}$ can change their causal character as well. In
this case, this has nothing to do with a horizon but it is also of
some relevance.

\begin{itemize}
	\item[(e)] $g_{tt} >0$. 
	If $a \neq 0$ the Killing field $\partial_t$ becomes spacelike,
	i.e. $g_{tt}=g(\partial _t , \partial _t)$ becomes positive, on part
	of the space-time. In this region an observer
	cannot move on a $t$-line. The region where $g_{tt} >0$ is known
	as the \emph{ergosphere} or the \emph{ergoregion}. (Note that some
	authors reserve this name for the intersection of the region where
	$g_{tt}>0$ with the domain of outer communication.)  

	\item[(f)] $g_{\varphi \varphi} <0$. 
	If $a \neq 0$ or $\ell \neq 0$, there is a region where the Killing
	field $\partial_{\varphi}$ becomes timelike. In this region, the
	space-time violates the causality condition because the
	$\varphi$-lines are closed timelike curves. If $\ell \neq 0$ and
	$\Lambda \le 0$, the region where this occurs extends to $r= \infty$.
	If $\ell \neq 0$ and $\Lambda > 0$, it is bounded by the first 
	(cosmological) horizon.
\end{itemize}

\section{Photon Regions}\label{sec:pregions}
In the space-times \eqref{eq:Metric}, the geodesic equation is 
completely integrable, i.e., it admits four constants of motion
in involution. These constants of motion are the Lagrangian
\begin{align}
	\Lag &= \tfrac{1}{2} g_{\mu\nu}\dot{x}^{\mu}\dot{x}^{\nu},
	\label{eq:L}
\intertext{the energy}
	E :&= -\frac{\partial\Lag}{\partial\dot{t}} 
		= -g_{\varphi t}\dot{\varphi} -g_{tt}\dot{t},
	\label{eq:E}
\intertext{the $z$-component of the angular momentum}
	L_{z} :&= \frac{\partial\Lag}{\partial\dot{\varphi}}
		= g_{\varphi\varphi}\dot{\varphi} +g_{\varphi t}\dot{t},
	\label{eq:Lz}
\end{align}
and the Carter constant $K$ \cite{Carter.1968b}. With the help of 
these four constants of motion, the geodesic equation can be 
written in first-order form. For lightlike geodesics, $\Lag=0$,
the resulting equations read
\begin{subequations}\label{eq:EoM}
	\begin{alignat}{1}
	\dot{t} &=
	\frac{\chi(L_{z}-E\chi)}{\Sigma\Delta_{\vartheta} \sin^{2}\vartheta}
		+\frac{(\Sigma + a\chi) \bigl((\Sigma + a\chi)E - aL_{z}\bigr)}{\Sigma\Delta_{r}},
\label{eq:EoM_t} \\
	\dot{\varphi}
	&=\frac{L_{z}-E\chi}{\Sigma\Delta_{\vartheta} \sin^{2}\vartheta}
		+\frac{a\bigl((\Sigma + a\chi)E - aL_{z}\bigr)}{\Sigma\Delta_{r}},
\label{eq:EoM_phi} \\
	\Sigma^{2} \dot{\vartheta}^{2} 
		&= \Delta_{\vartheta}K - \frac{(\chi E - L_{z})^{2}}{\sin^{2}\vartheta} 
		=: \Theta(\vartheta),
\label{eq:EoM_theta} \\
	\Sigma^{2} \dot{r}^{2} 
		&= \bigl((\Sigma + a\chi)E-aL_{z}\bigr)^{2} - \Delta_{r}K
		=: R(r).
\label{eq:EoM_r}
\end{alignat}
\end{subequations}
These equations can be solved explicitly in terms of hyperelliptic 
functions, see Hackmann et al. \cite{HackmannKagramanova.2009a}. 
Here, we are interested in spherical lightlike geodesics, i.e., lightlike
geodesics that stay on a sphere $r = \mathrm{constant}$. The region 
filled by these geodesics is called the \emph{photon region} $\K$.
To determine this photon region, we introduce the abbreviations
\begin{equation}\label{eq:LEKE}
	L_{E}=\frac{L_{z}}{E}, \quad K_{E}=\frac{K}{E^{2}}.
\end{equation}
For spherical orbits the conditions
$\dot{r}=0$ and $\ddot{r}=0$ have to be fulfilled. By 
\eqref{eq:EoM_r}, this requires that $R(r)=0$ and 
$R'(r)=0$, hence
\begin{equation}
\begin{aligned}
	K_{E} &= \frac{\bigl((\Sigma + a\chi)-aL_{E}\bigr)^{2}}{\Delta_{r}} ,  \\
	K_{E} &= \frac{4r\bigl((\Sigma + a\chi)-aL_{E}\bigr)}{\Delta_{r}'} ,
\end{aligned}
	\label{eq:2xKE}
\end{equation}
where $\Delta_{r}'$ denotes the derivative of $\Delta _r$ with respect
to $r$.
Solving for the constants of motion $K_{E}$ and $L_{E}$ results	in
\begin{align}
	K_{E} &= \frac{16r^{2}\Delta_{r}}{(\Delta_{r}')^{2}}, &
	aL_{E} &= \bigl(\Sigma + a\chi\bigr) - \frac{4r\Delta_{r}}{\Delta_{r}'}.
\label{eq:KL_SphLR}
\end{align}

Inserting these expressions into \eqref{eq:EoM_theta} and observing
that the left-hand side of \eqref{eq:EoM_theta} is non-negative
gives us an inequality that determines the photon region
\begin{equation}
	\K\colon \bigl(4r\Delta_{r} - \Sigma \Delta_{r}' \bigr)^{2}
		\leq 16 a^{2} r^{2} \Delta_{r} \Delta_{\vartheta} \sin^{2}\vartheta.
\label{eq:regionK}
\end{equation}
Note that $\K$ is independent of the Manko--Ruiz parameter $C$.
\pagebreak[1]

As in the Kerr case \citep[cf.][]{Perlick.2004}, through every point 
with coordinates ($r_{p}, \vartheta_{p}$) of $\K$ there is a 
lightlike geodesic which stays on the sphere $r=r_{p}$. Along each
of these spherical lightlike geodesics, the $\vartheta$ coordinate
oscillates between extremal values that are determined
by the equality sign in \eqref{eq:regionK}. The $\varphi$-motion is 
given by \eqref{eq:EoM_phi} and might be quite complicated. For
some spherical light rays it is not even monotonic.

In the non-rotating case ($a=0$) the inequality \eqref{eq:regionK}
degenerates into an equality,
\begin{equation}
	4r\Delta_{r} = (r^2+\ell ^2 ) \Delta_{r}'.
\label{eq:ps}
\end{equation}
This means that the photon regions degenerate into 
\emph{photon spheres}. The best known example is the
photon sphere in the Schwarzschild space-time at $r=3m$.

A spherical lightlike geodesic at $r=r_p$
is unstable with respect to radial 
perturbations if $R''(r_{p})>0$, and stable if $R''(r_{p})<0$. 
The second derivative  $R''$ can be calculated from 
\eqref{eq:EoM_r}. With the help of \eqref{eq:KL_SphLR}
this results in
\begin{equation}
	\frac{R''(r)}{8 E^{2}} \Delta_{r}'^{2} = 
		2r\Delta_{r}\Delta_{r}' + r^{2} \Delta_{r}'^{2} 
		- 2r^{2} \Delta_{r} \Delta_{r}''.
	\label{eq:stability}
\end{equation}

\begin{figure}[tbp]
	\centering
	\begin{tabular}{cl}
		\rectc{blau1}  & region with $\Delta_{r} \leq 0$  \\
		\rectc{rot}    & unstable spherical light-rays in $\K$  \\
		\rectc{orange} & stable spherical light-rays in $\K$  \\
		\rectl{blau2}  & region with $g_{\varphi \varphi}<0$ (causality violation)  \\
		\rectr{gray}   & region with $g_{tt}>0$ (ergosphere)  \\
		\circpdash{black,dashed}  & throats at $r=0$  \\
		\textbullet    & ring singularity
	\end{tabular}
	\caption{Legend for Figs. \ref{fig:Photonregion1}, \ref{fig:Photonregion2},
	\ref{fig:Photonregion3} and \ref{fig:PhotonregionCV}}
	\label{Legend}
\end{figure}

\begin{figure}[htbp]
	\centering
	\begin{tabular}{ll}
		$a=\frac{1}{50} a_{\mathrm{max}}$ &	\\
		\includegraphics{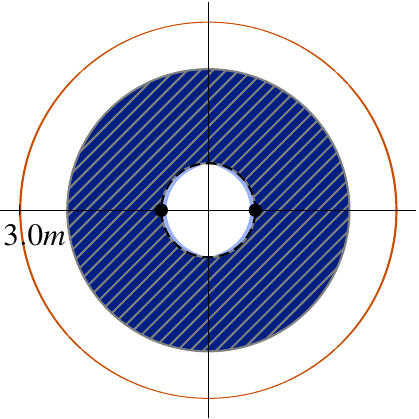} &
		\includegraphics[scale=0.6]{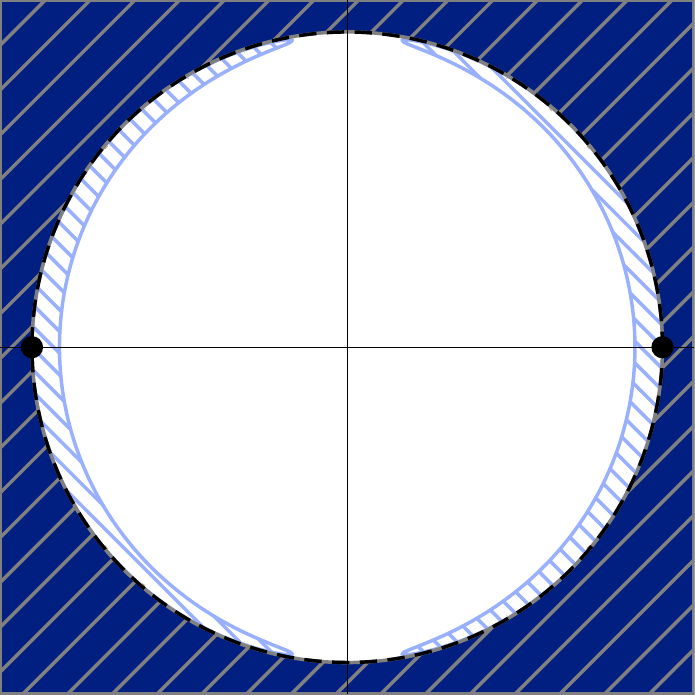} \\
\hline%
		$a=\frac{2}{5} a_{\mathrm{max}}$ &	\\
		\includegraphics{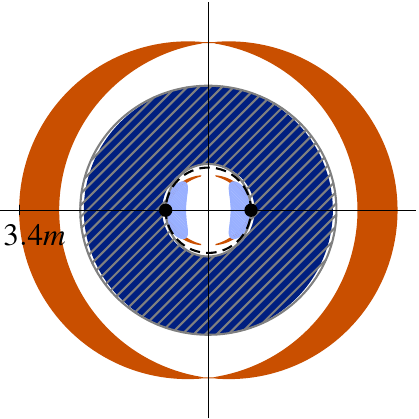} & 
		\includegraphics[scale=0.6]{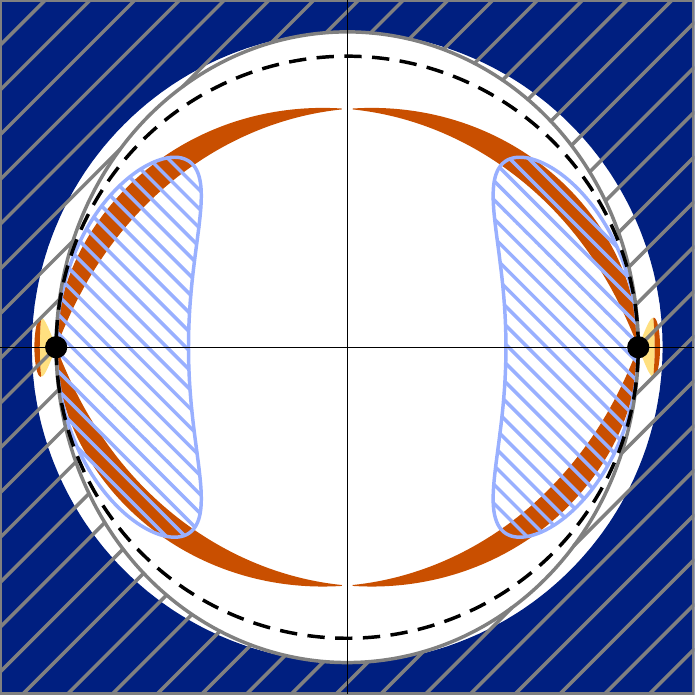} \\
\hline%
		$a=\frac{4}{5} a_{\mathrm{max}}$ &	\\
		\includegraphics{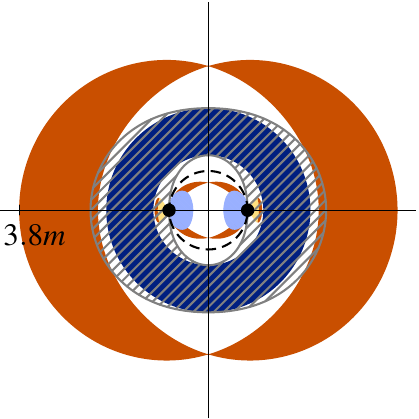} & 
		\includegraphics[scale=0.6]{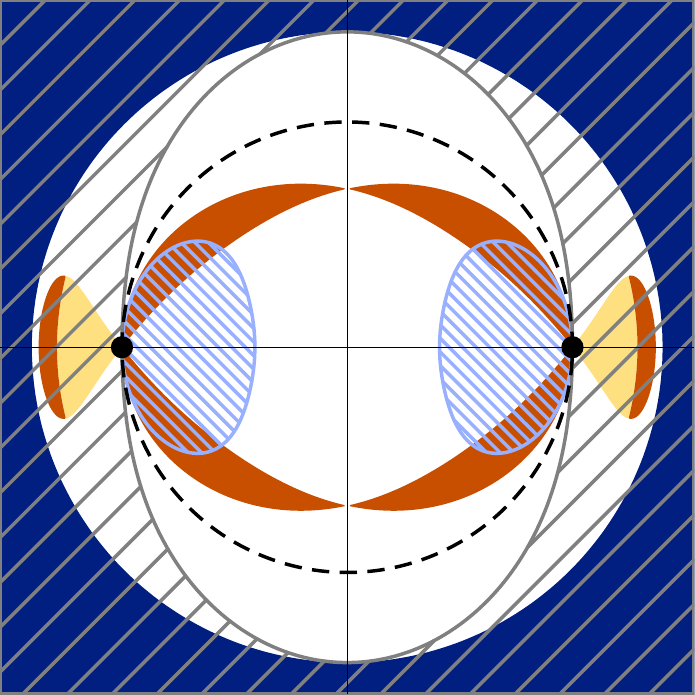} \\
\hline%
		$a=a_{\mathrm{max}}$ &	\\
		\includegraphics{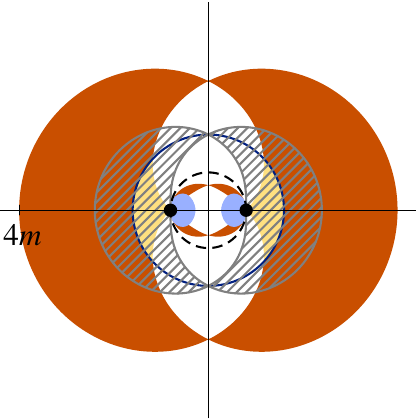} & 
		\includegraphics[scale=0.6]{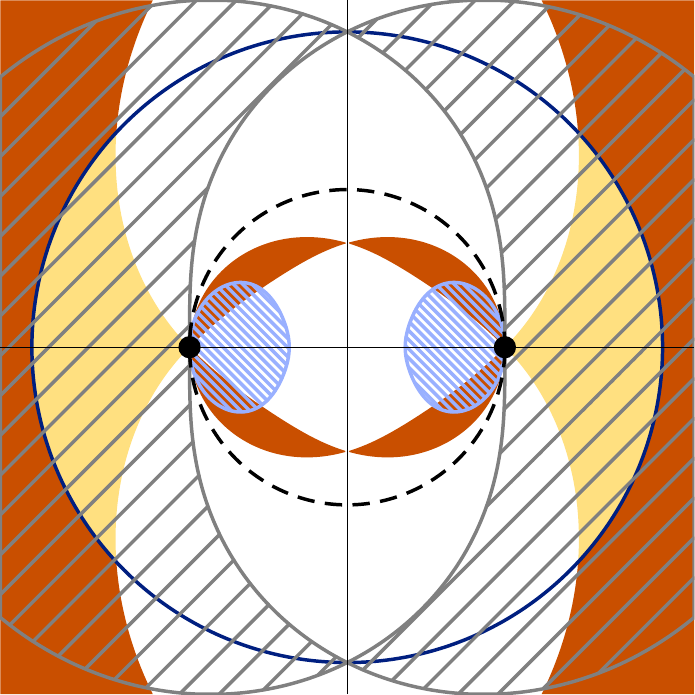} \\
	\end{tabular}
	\caption[Photon regions in Kerr space-time.]{%
	Photon regions in Kerr space-time for
	spins $a=\lambda a_{\mathrm{max}}$, where 
	$a_{\mathrm{max}}=m$. %
	The plots on the right show a magnified inner part.}
	\label{fig:Photonregion1}
\end{figure}

\begin{figure}[!tbp]
	\centering
	\begin{tabular}{ll}
		$a=\frac{1}{50} a_{\mathrm{max}}$ &	\\
		\includegraphics{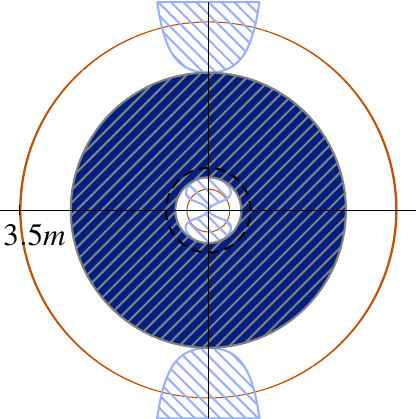} & 
		\includegraphics[scale=0.6]{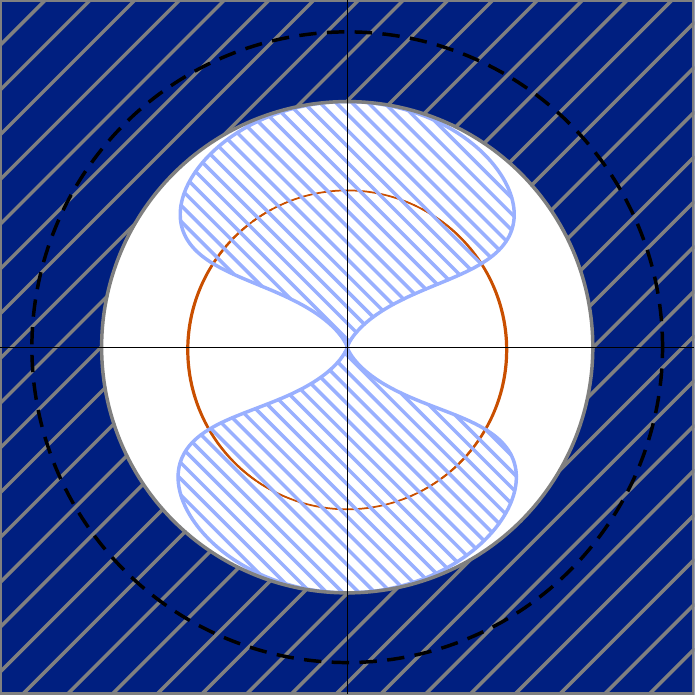} \\
\hline%
		$a=\frac{2}{5} a_{\mathrm{max}}$ &	\\
		\includegraphics{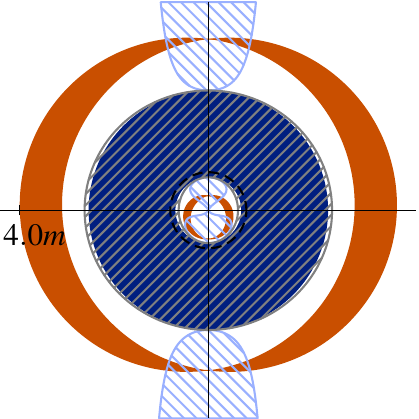} & 
		\includegraphics[scale=0.6]{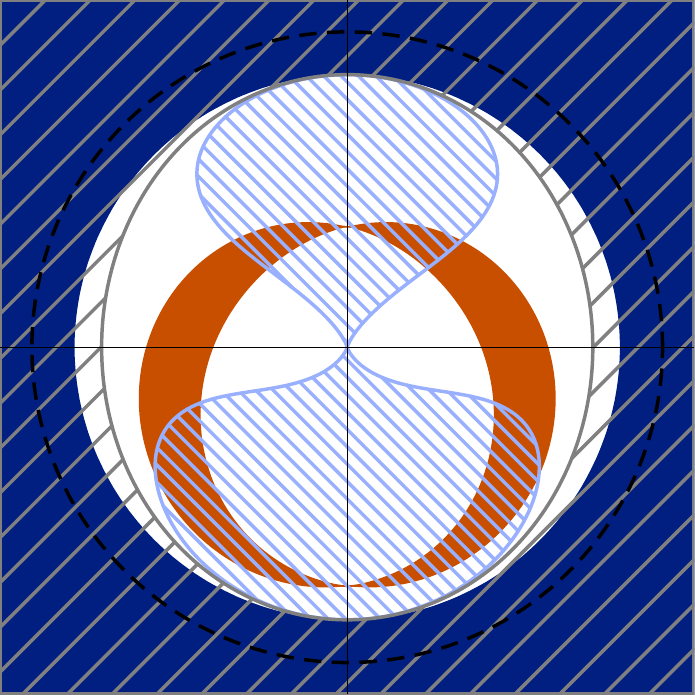} \\
\hline%
		$a=\frac{4}{5} a_{\mathrm{max}}$ &	\\
		\includegraphics{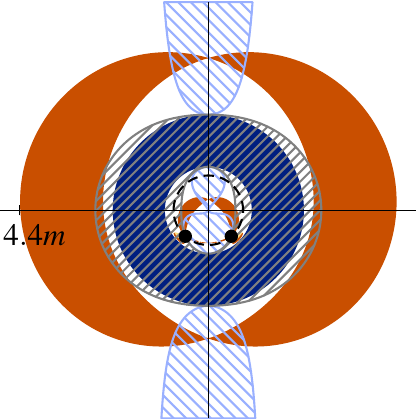} & 
		\includegraphics[scale=0.6]{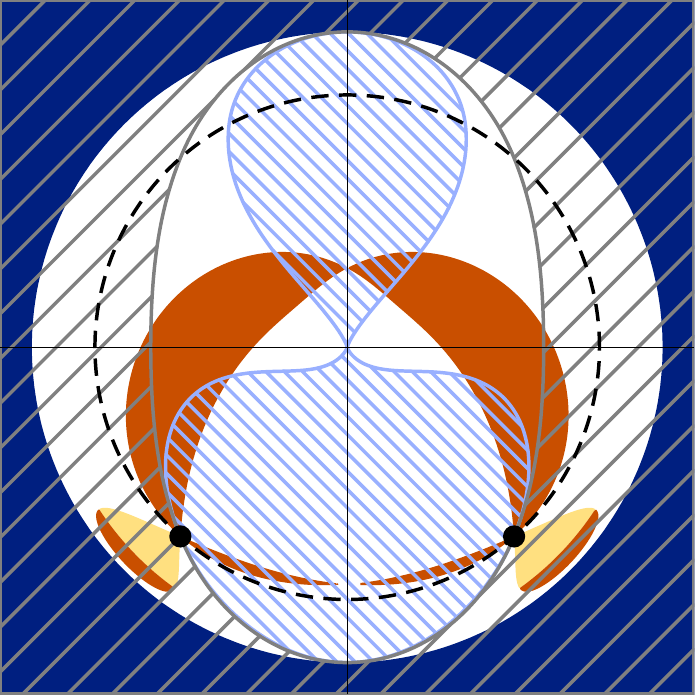} \\
\hline%
		$a=a_{\mathrm{max}}$ &	\\
		\includegraphics{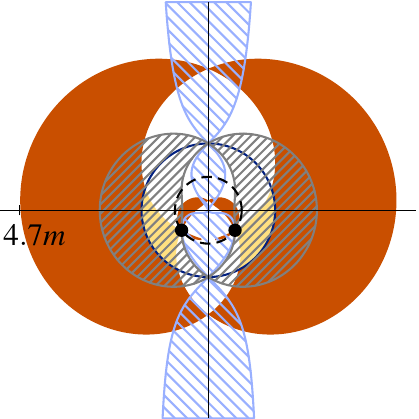} & 
		\includegraphics[scale=0.6]{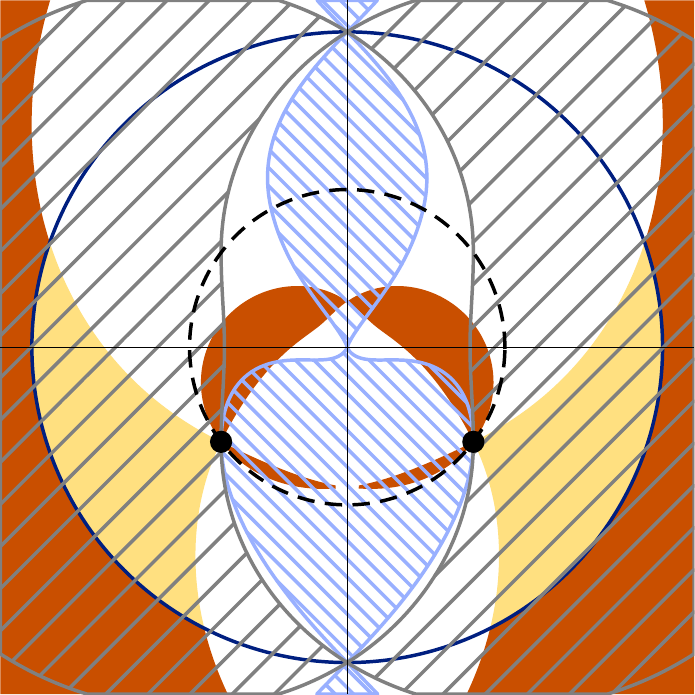} \\
	\end{tabular}
	\caption[Photon regions in Kerr--NUT space-time.]{%
	Photon regions in Kerr--NUT space-time with 
	$\ell=\frac{3}{4}m$, $C=0$ for spins 
	$a=\lambda a_{\mathrm{max}}$, where 
	$a_{\mathrm{max}}=\sqrt{m^{2}+\ell^{2}}=\tfrac{5}{4}m$. %
	The plots on the right show a magnified inner part.}
	\label{fig:Photonregion2}
\end{figure}

\begin{figure}[!tbp]
	\centering
	\begin{tabular}{ll}
		$a=\frac{1}{50} a_{\mathrm{max}}$ &	\\
		\includegraphics{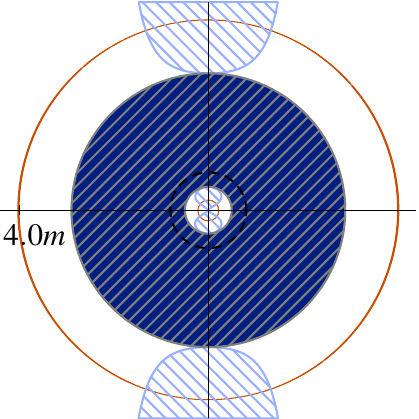} & 
		\includegraphics[scale=0.6]{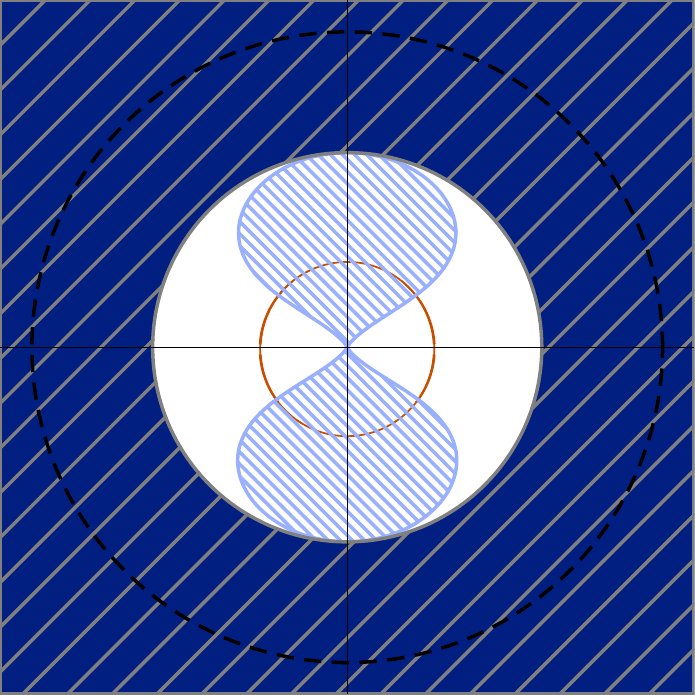} \\
\hline%
		$a=\frac{2}{5} a_{\mathrm{max}}$ &	\\
		\includegraphics{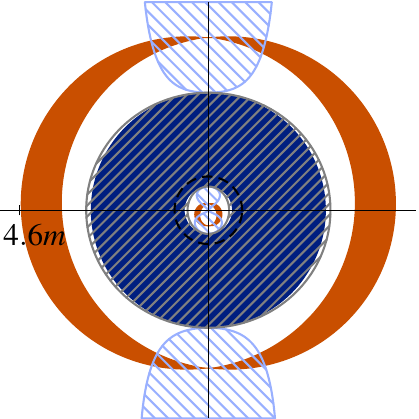} & 
		\includegraphics[scale=0.6]{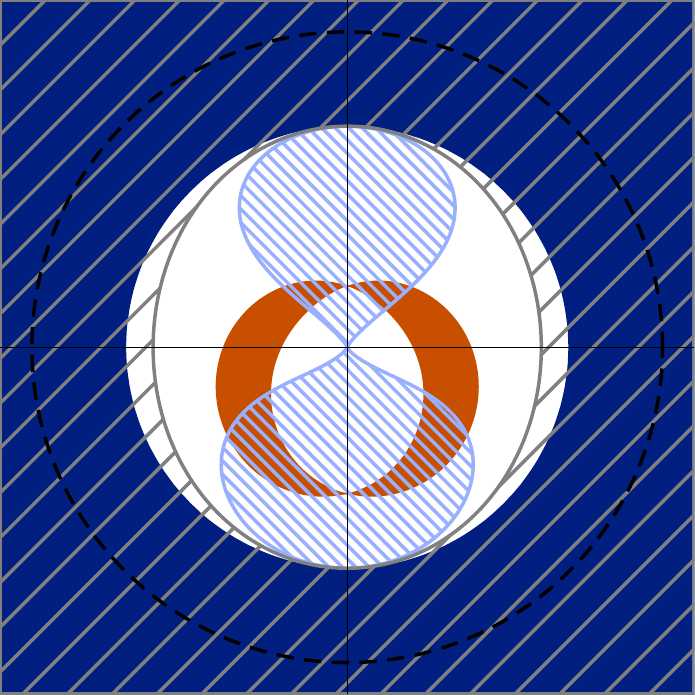} \\
\hline%
		$a=\frac{4}{5} a_{\mathrm{max}}$ &	\\
		\includegraphics{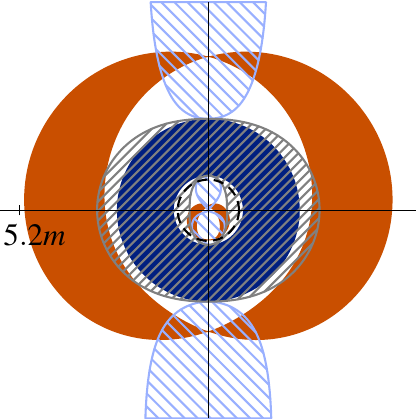} & 
		\includegraphics[scale=0.6]{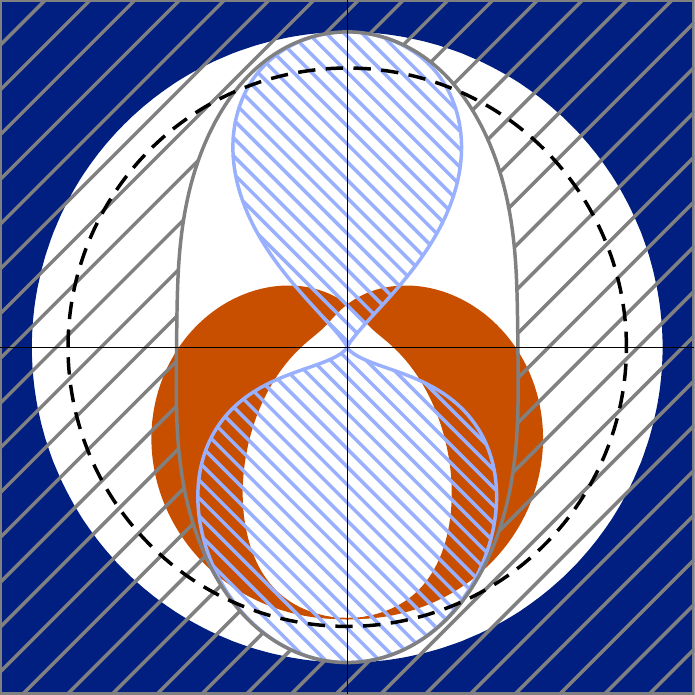} \\
\hline%
		$a= a_{\mathrm{max}}$ &	\\
		\includegraphics{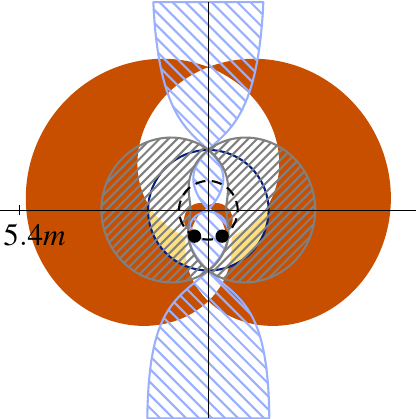} & 
		\includegraphics[scale=0.6]{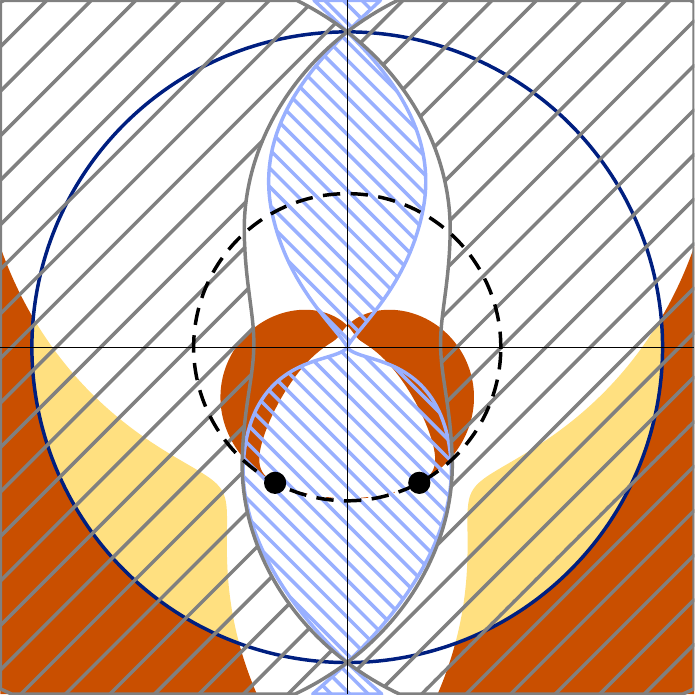} \\
	\end{tabular}
	\caption[Photon regions in Kerr--Newman--NUT
	space-time with cosmological constant.]{Photon
	regions in Kerr--Newman--NUT space-time ($\beta=\frac{5}{9}m^{2}$,
	$\ell=\frac{4}{3}m$, $C=0$) with a cosmological constant
	($\Lambda={10}^{-2}m^{-2}$) for spins 
	$a=\lambda a_{\mathrm{max}}$, where 
	$a_{\mathrm{max}}\approx 1.51\,m$. %
	The plots on the right show a magnified inner part.}
	\label{fig:Photonregion3}
\end{figure}

\begin{figure*}[htbp]
	\setlength{\tabcolsep}{5pt}
	\centering
	\begin{tabular}{ccccccc}
		\includegraphics{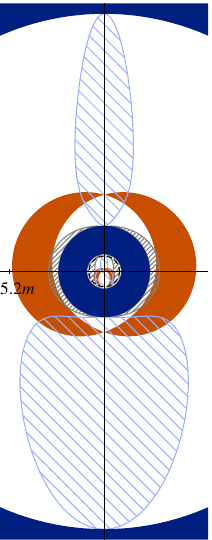} & 
		\includegraphics{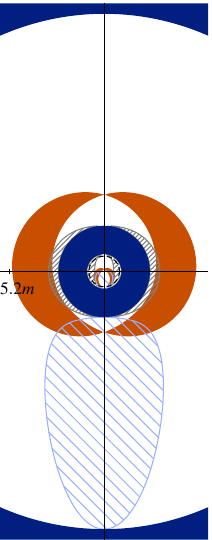} & 
		\includegraphics{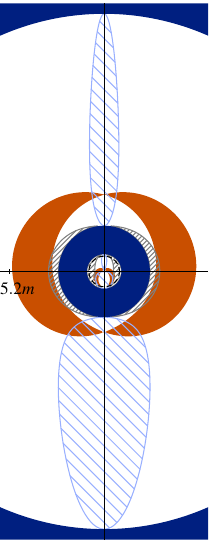} & 
		\includegraphics{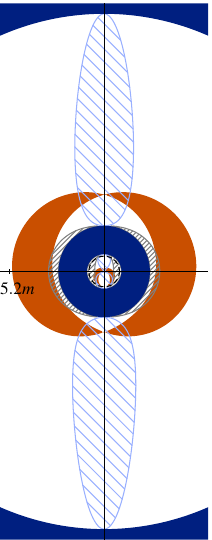} & 
		\includegraphics{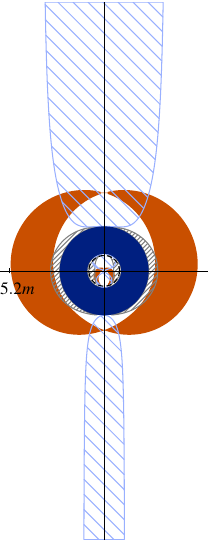} & 
		\includegraphics{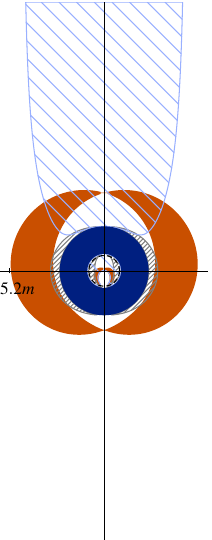} & 
		\includegraphics{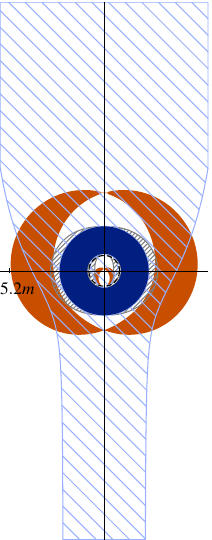} \\
		$C=-2$ & $C=-1$ & $C=-\frac{1}{2}$ & $C=0$ & 
		$C=\frac{1}{2}$ & $C=1$ & $C=2$ 
	\end{tabular}
	\caption[Photon regions for varying singularity 
	parameter $C$ with different spins $a$ and fixed $\beta$, 
	$\ell$.]{Photon regions for varying singularity 
	parameter $C$ with fixed $a=\frac{4}{5} a_{\mathrm{max}}$, 
	$\beta=\frac{5}{9}m^{2}$, $\ell=\frac{4}{3}m$, and 
	$\Lambda = \bigl\{\begin{smallmatrix}
		{10}^{-2}m^{-2} & \text{for } C\leq 0  \\
		0 & \text{for } C>0
	\end{smallmatrix}\bigr.$,
	where
	$a_{\mathrm{max}} = \bigl\{\begin{smallmatrix}
		1.51m & \text{for } C\leq 0  \\
		2\sqrt{5}\,m/3 & \text{for } C>0
	\end{smallmatrix}\bigr.$.
	If existent, the cosmological horizon restricts the region
	(\rectl{blau2}) where the causality is violated. If $C=1$ or
	$C=-1$, one of the two half-axes is regular and it is not
	surrounded by a causality violating region.}
	\label{fig:PhotonregionCV}
\end{figure*}

Figs.~\ref{fig:Photonregion1}, \ref{fig:Photonregion2},
\ref{fig:Photonregion3} and \ref{fig:PhotonregionCV} 
show plots of the photon region $\K$ in the $(r,\vartheta)$ 
plane, where unstable (\rectc{rot}) and stable
(\rectc{orange}) spherical light rays \eqref{eq:stability} are
distinguished. The boundaries of the region where $\Delta_{r}\leq 0$ 
(\rectc{blau1}) are the horizons.
Furthermore, the ergosphere (\rectr{gray}), the causality violating
region (\rectl{blau2}), and the ring singularity (\textbullet) are
shown. A legend for these figures can be found in Fig.~\ref{Legend}.

Each picture illustrates a meridional section through space-time, 
i.e. the plane parametrized by $r$ and $\vartheta$, where the 
$\vartheta$-coor\-dinate is measured from the positive $z$-axis. 
Following a suggestion by O'Neill \cite{ONeill.1995}, 
we show the whole range of the space-time, with the Boyer--Lindquist
coordinate $r$ increasing outward from the origin which corresponds to 
$r= -\infty$. O'Neill suggested to use the exponential of $r$ for the
radial coordinate. As such a representation strongly exaggerates the
outer parts, we find it more convenient to use two different scales.
In the region $r<0$ (i.e., inside the sphere $r=0$), we use 
$m \exp \big( r/m \big)$ for the radial coordinate.  In the region
$r>0$ (i.e., outside the sphere $r=0$), we use $r+m$ for the radial
coordinate.  The dashed circle (\circpdash{black}) indicates the
throats at $r=0$.

Each figure shows the photon region for four different values of 
the spin $a$, keeping all the other parameters fixed. Restricting 
to black-hole cases, we choose the four values of the spin as 
$a=\lambda a_{\mathrm{max}}$, where
$\lambda\in\bigl\{\frac{1}{50}, \frac{2}{5}, \frac{4}{5}, 1\bigr\}$
and $a_{\mathrm{max}}$ denotes the spin of 
an extremal black hole which is determined by the other 
parameters. If $\Lambda =0$, we have $a_{\mathrm{max}}^{2}=
m^{2}+\ell^{2}-\beta$, cf. Eq.~\eqref{eq:InOutHorizon}. 
If $\Lambda \neq 0$, there is no convenient formula for
$a_{\mathrm{max}}$ because one has to evaluate a fourth-order
equation.

In the Kerr space-time, see Fig.~\ref{fig:Photonregion1}, there
is an exterior photon region at $r>r_+$ and an interior photon
region at $r<r_-$. Both of them are 
symmetric with respect to the equatorial plane. Starting 
from the photon sphere at $r=3m$ for the non-rotating 
Schwarzschild case, the exterior photon region gets a 
crescent-shaped cross-section for $a\neq 0$ and grows with increasing
spin $a$. The interior photon region consists of two connected
components that are separated by the ring singularity.
In the exterior photon region all spherical light orbits 
are unstable while in the interior photon region there are stable
and unstable ones. Circular lightlike geodesics exist where the 
boundary of the photon region is tangent to a sphere $r= \mathrm{constant}$.
We easily recognize the three well-known circular lightlike geodesics 
in the equatorial plane, but also two not-so-well-known cicular
lightlike geodesics off the equatorial plane. The latter are
situated in the region where $r<0$. The causality
violating region is adjacent to the ring singularity and lies
to the side of negative $r$. For small $a$,
the ergoregion does not intersect the exterior photon region
but for $a^2 > m^2/2$ it does.

The additional gravitomagnetic charge $\ell$ of the Kerr--NUT
space-time changes the symmetry behavior significantly, see
Fig.~\ref{fig:Photonregion2}. The plots are no longer symmetric
with respect to the equatorial plane (but they remain, of course,
axially symmetric). The exterior and interior photon regions
show this asymmetry clearly.  For a slowly rotating 
Kerr--NUT black hole, $a^2< \ell ^2$, there is no ring singularity, 
and there are no stable spherical light rays. If the spin is increased, 
the ring singularity appears at $a^2= \ell ^2$, degenerated to a point 
on the axis. With $a$ further increased, the ring singularity moves
towards the equator and stable spherical light orbits come into 
existence between $r=0$ and $r=r_-$; as in the Kerr case, the
interior photon region consists of two connected components
that are separated by the ring singularity. While the ergosphere 
is not significantly affected by $\ell$, there is an additional 
causality violating region around the singularity on the
axis which extends from the outer horizon at $r=r_+$ to 
$r=\infty$. The interior causality violating region is now
extending from the inner horizon at $r=r_-$ to $r=-\infty$. 
The causality violating region depends on the Manko-Ruiz parameter 
$C$ which was chosen equal to zero in Fig.~\ref{fig:Photonregion2}. 
(For other values of $C$ see Fig.~\ref{fig:PhotonregionCV}.)

Adding an electric or magnetic charge parameter $\beta$ and a
cosmological constant $\Lambda$ affects the photon regions little, 
see Fig.~\ref{fig:Photonregion3}. The only qualitative effect
of $\beta$ is in the fact that, in the case $a^2 > \ell ^2$, 
one of the two connected components of the interior photon 
region is now detached from the ring singularity. For non-zero
$\Lambda$, higher spin values $a_{\mathrm{max}}$ are possible 
compared to space-times with $\Lambda=0$. For the pictures we 
have chosen a (small and) positive value for $\Lambda$ such that 
the domain of outer communication is bounded by a cosmological
horizon. The latter is not shown in Fig.~\ref{fig:Photonregion3}
because these pictures do not extend so far, but it is shown in 
Fig.~\ref{fig:PhotonregionCV}. The cosmological horizon  
restricts the causality violating region which depends on the 
Manko-Ruiz parameter $C$, see Fig.~\ref{fig:PhotonregionCV}.

\section{Shadows of Black Holes}\label{sec:shadow}
The existence of the photon region \eqref{eq:regionK} around the black
hole is essential for the construction of the \emph{shadow} of a black 
hole. 
In the Introduction we have already explained how
the shadow is constructed in the case of a Schwarzschild black hole.
The same construction works, mutatis mutandis, in our more general
black-hole space-times. We fix an observer in the domain of outer 
communication at Boyer-Lindquist coordinates $(r _O , \vartheta _O)$ 
and we think of light sources distributed on a sphere $r=r_L$ with some 
$r_L>r_O$.

For determining the shape of the shadow it is convenient to consider
light rays which are sent from the observer's position \emph{into the
past}. Then we can distinguish two types of orbits. Along light rays
of the first type the radius coordinate reaches the value 
$r_L$, possibly after going through a local minimum, so that we can 
think of these light rays as being emitted from one of our light sources. 
Along light rays of the second type the radius coordinate decreases 
monotonically until it reaches the horizon at $r=r_+$, so these light 
rays cannot come from any of our light sources. Correspondingly, in the 
direction of light rays of the first type the observer would see brightness, 
and in the direction of light rays of the second type the observer would 
see darkness. The borderline case, i.e. the boundary of the shadow,
corresponds to light rays that asymptotically spiral towards one of
the unstable spherical light orbits in the exterior photon region
which was discussed in Section~\ref{sec:pregions} above. 
As in the Schwarzschild case, it is obvious from the geometry that 
the construction of the shadow works equally well if light sources
are distributed, rather than on a sphere $r=r_L$ with $r_L>r_O$, anywhere 
else in the domain of outer communication except in the region filled
by the above-mentioned light rays of the second type.

\begin{figure}[t]
	\centering
	\includegraphics{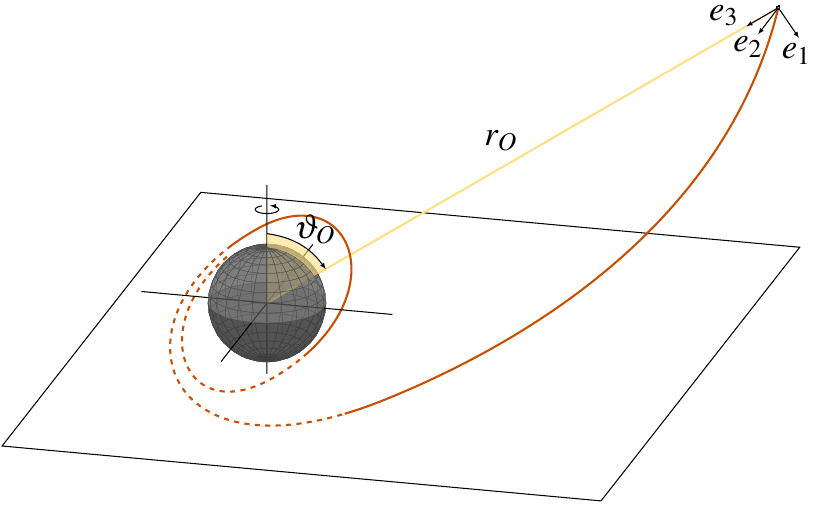}
	\caption{At an observation event with Boyer--Lindquist coordinates
	$(r_{O},\vartheta_{O})$ we choose an orthonormal tetrad
	$(e_0,e_1,e_2,e_3)$ according to Eqs.~\eqref{eq:newcoord}. 
	For each light ray that is sent from the observation event into 
	the past the tangent vector can be written as a linear combination of 
	$e_0$, $e_1$, $e_2$ and $e_3$. In this way we can assign celestial
	coordinates to the direction of the tangent vector, see 
	Fig.~\ref{fig:LightObserv}}
	\label{fig:CoordCh}
\end{figure}

It is now our goal to calculate the boundary curve of the shadow on
the observer's sky. We consider an observer at position
$(r_{O},\vartheta_{O})$ in the Boyer--Lindquist coordinates. (The
$\varphi$ and $t$ coordinates of the observation event are irrelevant
because of the symmetries of the metric.) We choose an orthonormal
tetrad
\begin{equation}
\begin{aligned}
	e_{0} &= \left.
		\frac{(\Sigma + a \chi) \partial_t + a \partial_{\varphi}}{
		\sqrt{\Sigma \Delta_r}}\right|_{(r_O,\vartheta_O)}, \\[\smallskipamount]
	e_{1} &= \left.
		\sqrt{\dfrac{\Delta _{\vartheta}}{\Sigma}} \, \partial_{\vartheta}
		\right|_{(r_O,\vartheta_O)}, \\[\smallskipamount]
	e_{2} &= \left.
		\frac{-(\partial_{\varphi} + \chi \partial_t)}{
		\sqrt{\Sigma \Delta_{\vartheta}}  \sin \vartheta} 
		\right|_{(r_O,\vartheta_O)}, \\[\smallskipamount]
	e_{3} &= \left.
		-\sqrt{\frac{\Delta_r}{\Sigma}} \, \partial_r
		\right|_{(r_O,\vartheta_O)}.
\end{aligned}
\label{eq:newcoord}
\end{equation}
at the observation event (see Fig.~\ref{fig:CoordCh}). We assume that 
the observer is in the domain of outer communication. This guarantees 
that $\Delta_r$ is positive, and so is $\Sigma$. Moreover, we assume 
that $\ell$ and $\Lambda$ are restricted by the inequality 
\eqref{eq:ellLambda}, which guarantees that $\Delta_{\vartheta}$ is 
positive. Hence, the coefficients in Eqs.~\eqref{eq:newcoord} are indeed 
real and it is straight-forward to verify that $e_0$, $e_1$, $e_2$, $e_3$ 
are orthonormal. The timelike vector $e_0$ is to be interpreted as the 
four-velocity of our observer. The tetrad has been chosen such that 
$e_0 \pm e_3$ are tangential to the \emph{principal null congruences} 
of our metric. For an observer with four-velocity $e_0$ the vector 
$e_3$ gives the spatial direction towards the center of the black hole.
  
\begin{figure}[htbp]
	\centering
	\includegraphics{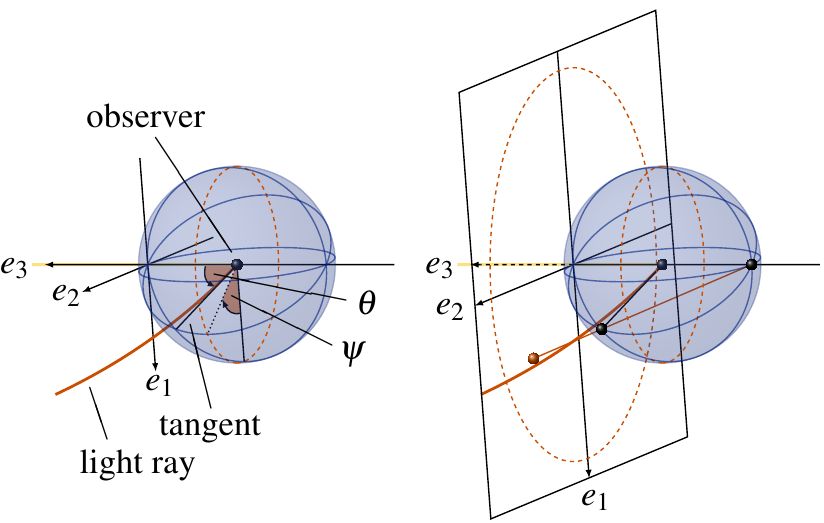}
	\caption{To each light ray at the observation event we assign
	celestial coordinates $\theta$ and $\psi$ with the help of
	Eq. \eqref{eq:dotlambda2}, see figure on the left. The figure 
	on the right shows the stereographic projection (red
	ball) of the point $(\theta,\psi)$ on the celestial sphere (black
	ball). The dotted (red) circles indicate the celestial equator 
	$\theta = \pi/2$ and its projection.}
	\label{fig:LightObserv}
\end{figure}

For each light ray $\lambda (s)$ with coordinate representation
$\bigl(r(s),\vartheta(s),\varphi(s),t(s)\bigr)$, we write the tangent
vector as
\begin{equation}
	\dot{\lambda} = \dot{r} \partial_{r} + \dot{\vartheta} \partial_{\vartheta}
	+ \dot{\varphi} \partial_{\varphi} + \dot{t} \partial_{t}.
\label{eq:dotlambda1}
\end{equation}
On the other hand, the tangent vector at the observation event
can be written as
\begin{equation}
	\dot{\lambda} = \alpha \big( -e_{0} + \sin\theta \cos\psi e_{1} 
	+ \sin\theta \sin\psi e_{2} + \cos\theta e_{3} \big) 
\label{eq:dotlambda2}
\end{equation}
where $\alpha$ is a scalar factor. From \eqref{eq:E} and 
\eqref{eq:Lz} we find that
\begin{equation}
	\alpha = \left.
		g \big(\dot{\lambda},e_0 \big) =
		\frac{a L_z - (\Sigma+a\chi)E}{\sqrt{\Sigma \Delta_r}} 
		\right|_{(r_O,\vartheta_O)}.
\label{eq:alpha}
\end{equation}
Eq.~\eqref{eq:dotlambda2} defines the celestial coordinates $\theta$
and $\psi$ for our observer, see Fig.~\ref{fig:LightObserv}.  
The direction towards the black hole corresponds to $\theta =0$.

Comparing coefficients of $\partial_{\varphi}$ and $\partial_r$ in 
\eqref{eq:dotlambda1} and \eqref{eq:dotlambda2} yields
\begin{equation}
\begin{aligned}
	\sin\psi &= \left.
		\frac{\sqrt{\Delta_{\vartheta}} \sin \vartheta}{\sqrt{\Delta_r} \sin\theta } 
		\left( \frac{\Sigma \Delta_r \dot{\varphi}}{
			(\Sigma + a \chi)E - aL_z} - a\right) 
		\right|_{(r_O,\vartheta_O)}, \\[\smallskipamount]
	\cos\theta &= \left.
		\frac{\Sigma  \dot{r}}{(\Sigma + a\chi)E - aL_z} 
	\right|_{(r_O,\vartheta_O)}.
\end{aligned}
\label{eq:relation_theta_psi}
\end{equation}
Upon substituting for $\dot{\varphi}$ and $\dot{r}$ from
\eqref{eq:EoM_phi} and \eqref{eq:EoM_r} we find from
\eqref{eq:relation_theta_psi} that
\begin{equation}
\begin{aligned}
	\sin\psi &= \left.
		\frac{\widetilde{L}_E + a\cos^2\vartheta + 2\ell \cos\vartheta}{
			\sqrt{\Delta_{\vartheta} K_E} \sin\vartheta} 
		\right|_{\vartheta = \vartheta_O}, 
	\\[\smallskipamount]
	\sin\theta &= \left. 
		\frac{\sqrt{\Delta_r K_E}}{r^2 + \ell^2 - a\widetilde{L}_E} 
		\right|_{r=r_O},
\end{aligned}
\label{eq:psitheta2}
\end{equation}
where
\begin{equation}
	\widetilde{L}_E = L_E - a + 2 \ell C.
	\label{eq:tLE}
\end{equation}
The boundary curve of the shadow corresponds to light rays
that asymptotically approach a spherical lightlike geodesic.
Such a light ray must have the same constants of motion as 
the limiting spherical lightlike geodesic, i.e., 
by \eqref{eq:KL_SphLR},
\begin{equation}
\begin{aligned}
	K_{E} &= \left. \frac{16r^{2}\Delta_{r}}{(\Delta_{r}')^{2}} \right|_{r=r_p}, 
	\\[\smallskipamount]
	a\widetilde{L}_{E} &= \left. \Big(r^2 + \ell^2  - \frac{4r\Delta_{r}}{
		\Delta_{r}'} \Big)\right|_{r=r_p},
\end{aligned}
\label{eq:LEKEc}
\end{equation}
where $r_p$ is the radius coordinate of the limiting 
spherical lightlike geodesic. Inserting the expressions 
for $K_E$ and $\widetilde{L}_E$ from \eqref{eq:LEKEc} 
into \eqref{eq:psitheta2} gives the boundary curve 
$\big( \psi (r_p),\theta (r_p) \big)$ of the shadow.

We observe that the Manko-Ruiz parameter $C$ has no
influence on the shadow and that the shadow is always 
symmetric with respect to a horizontal axis. The 
latter result follows from the fact that the points 
$(\psi,\theta)$ and $(\pi - \psi, \theta)$ correspond
to the same constants of motion $K_E$ and 
$\widetilde{L}_E$. For $\ell \neq 0$ and $\vartheta _O
\neq \pi/2$ 
this symmetry property was not to be expected. 

For $a >0$, the $\theta$ coordinate takes its maximal
value along the boundary curve at $\psi = - \pi /2$
and its minimal value at $\psi = \pi/2$. The corresponding 
values of the parameter $r_p$, which we denote by 
$r_{\mathrm{max}}(\vartheta _O)$ and 
$r_{\mathrm{min}}(\vartheta _O)$, respectively,
can be determined by inserting \eqref{eq:LEKEc} into
\eqref{eq:psitheta2} and equating $\psi$ to $\mp \pi /2$.
We find that $r_p=r_{\mathrm{max/min}}(\vartheta _O)$ is determined
by the equation
\begin{equation}\label{eq:rmaxmin}
	\Big( \Sigma \Delta_r ' - 4r \Delta_r 
	\mp 4ar \sqrt{\Delta _r \Delta _{\vartheta}} \,  
	\sin \vartheta \Big) 
	\Big| _{(r=r_p, \vartheta = \vartheta _O)} = \, 0 \, .
\end{equation}
Comparison with the inequality \eqref{eq:regionK}
shows that $r_{\mathrm{max}}(\vartheta _O)$ and 
$r_{\mathrm{min}}(\vartheta _O)$ are the radius values 
where the boundary of the exterior photon region 
intersects the cone $\vartheta = \vartheta _O$.

The case $a =0$ is special because then our method of   
parametrizing the boundary curve  by $r_p$ does not work.
If $a=0$ we have $r_{\mathrm{min}} (\vartheta _O)= r_p 
= r_{\mathrm{max}}(\vartheta _O)$, 
so \eqref{eq:LEKEc} determines a unique value for $K_E$. Inserting
this value into \eqref{eq:psitheta2} gives the boundary curve of the
shadow in the form
$\big( \psi(\widetilde{L}_E), \theta(\widetilde{L}_E) \big)$. We see
that $\theta = \mathrm{constant}$ if $a=0$, i.e., that the
shadow is circular.

Note that we have calculated the shadow for an observer with
four-velocity $e_0$  according to \eqref{eq:newcoord}. For an 
observer with a different four-velocity the shadow is distorted 
according to the standard aberration formula of special relativity.
  
\begin{figure}[htbp]
	\centering
	\begin{tabular}{ll}
		$a=\frac{2}{5} a_{\mathrm{max}}$ & 
		$a=\frac{4}{5} a_{\mathrm{max}}$ \\
		\includegraphics[scale=0.5]{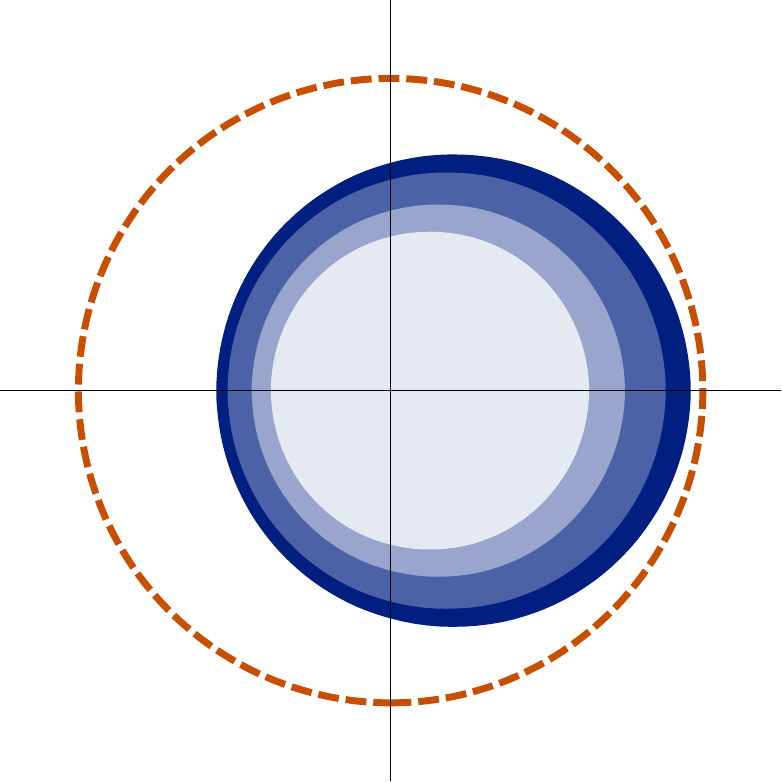} &
		\includegraphics[scale=0.5]{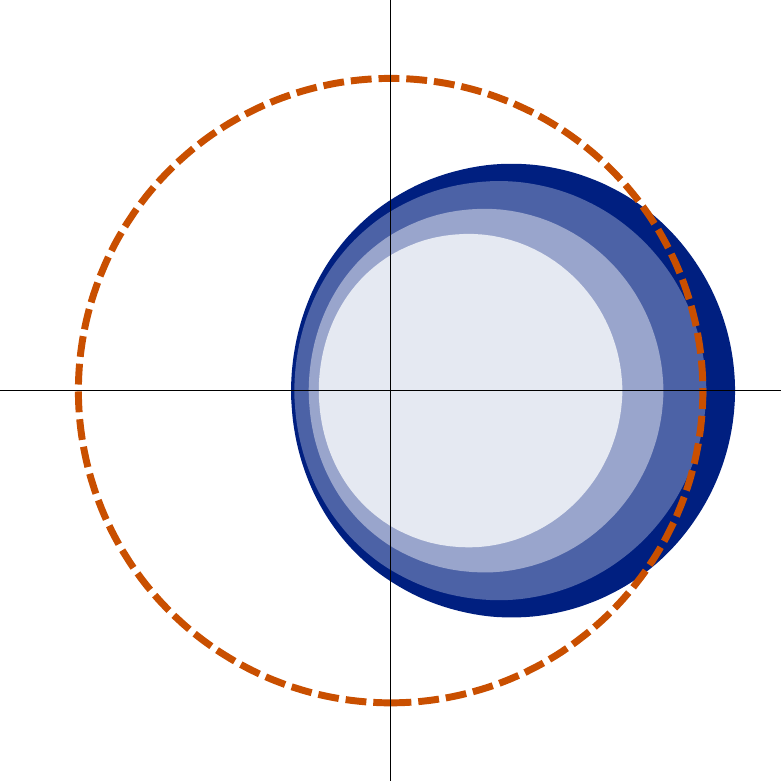} \\
		$a=\frac{1}{50} a_{\mathrm{max}}$ & 
		$a=a_{\mathrm{max}}$ \\
		\includegraphics[scale=0.5]{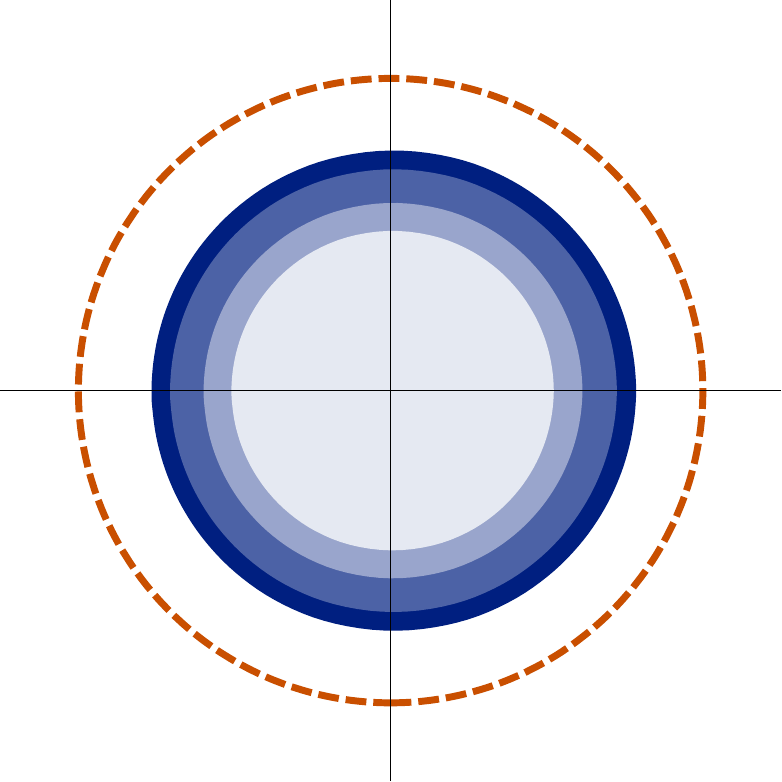} &
		\includegraphics[scale=0.5]{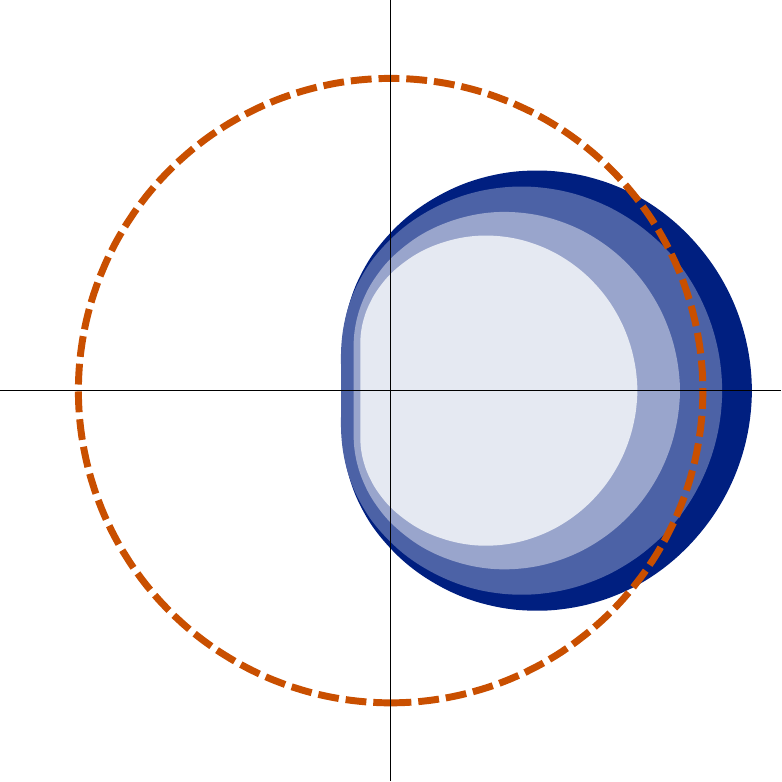} \\
	\end{tabular}
	\begin{tabular}{C|CCCCC}
		\multicolumn{1}{m{6mm}|}{} & \M{\circc{blau1!10!white}} & 
		\M{\circc{blau1!40!white}} & \M{\circc{blau1!70!white}} & 
		\M{\circc{blau1}} \\
	\hline
		\beta & 0 & 0 & \frac{5}{9}m^{2} & 0 \\
		\ell & 0 & \frac{3}{4}m & \frac{4}{3}m & \frac{4}{3}m \\
		\Lambda & 0 & 0 & {10}^{-2}m^{-2} & 0 \\
	\hline
		{a_{\mathrm{max}}} & {m} & {\frac{5}{4}m} & {1.51m} & {\frac{5}{3}m} \\[2pt]
	\hline
		& \M{Kerr} & \M{Kerr--NUT} & \M{KN--NUT with $\Lambda$} & \M{Kerr--NUT}
	\end{tabular}
	\caption[Shadows for different parameters $a$,
	$\beta$, $\ell$, $C$, and $\Lambda$.]{Shadow of a black hole 
	for different parameters $a$, $\beta$, $\ell$ and $\Lambda$, seen
	by an observer at $r_O=5m$ and $\vartheta_O = \pi /2$. The cross 
	hairs indicate the spatial direction towards the black hole, i.e., 
	the spatial direction of the principal null congruences with respect
	to our observer with four-velocity $e_0$. The dashed (red) circle 
	indicates the celestial equator, cf. Fig.~\ref{fig:LightObserv}.}
	\label{fig:Shadow}
\end{figure}

\begin{figure*}[htbp]
	\setlength{\tabcolsep}{5pt}
	\centering
	\begin{tabular}{ccccc}
		\includegraphics[scale=0.4]{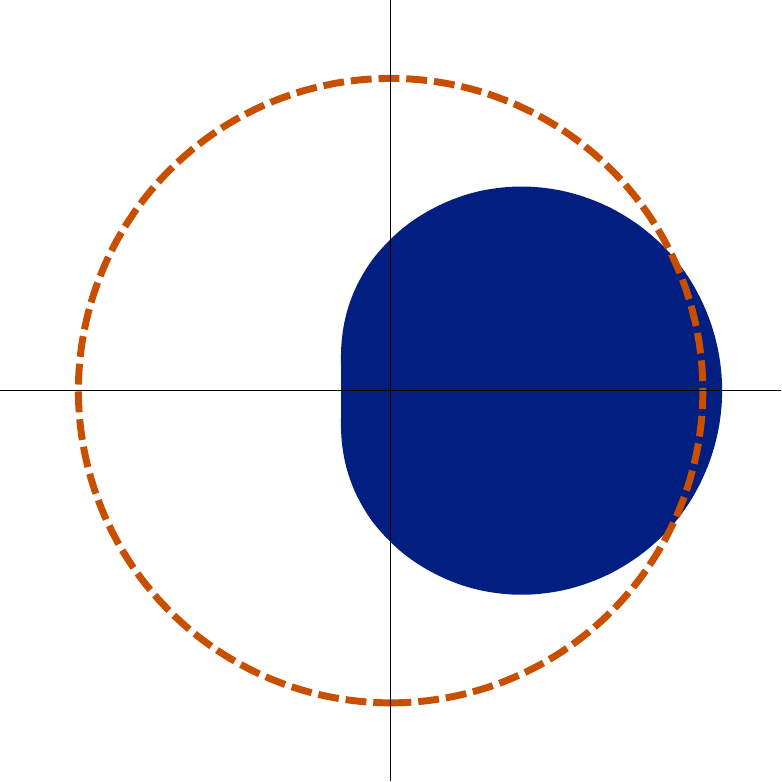}	& 
		\includegraphics[scale=0.4]{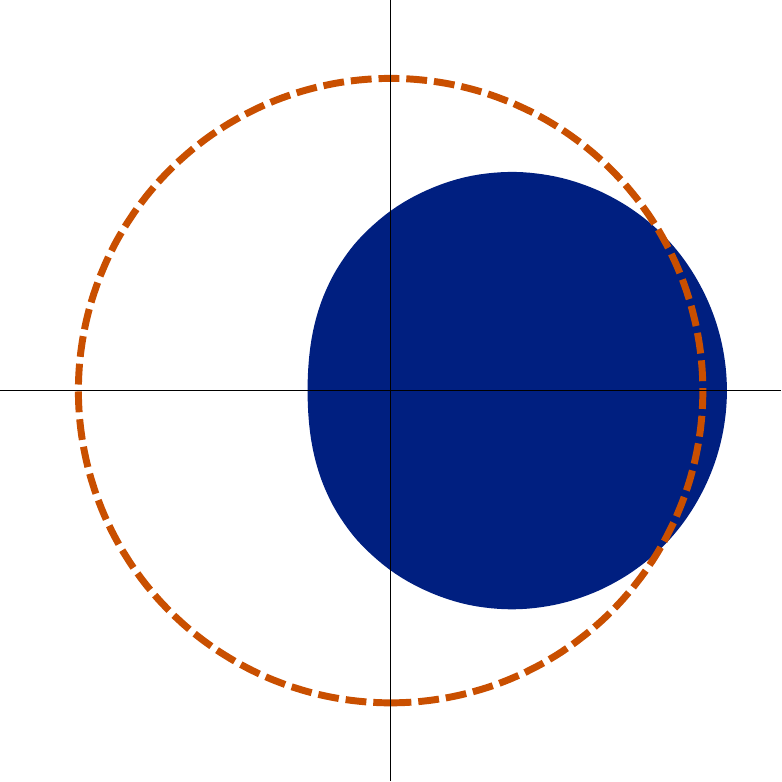}	& 
		\includegraphics[scale=0.4]{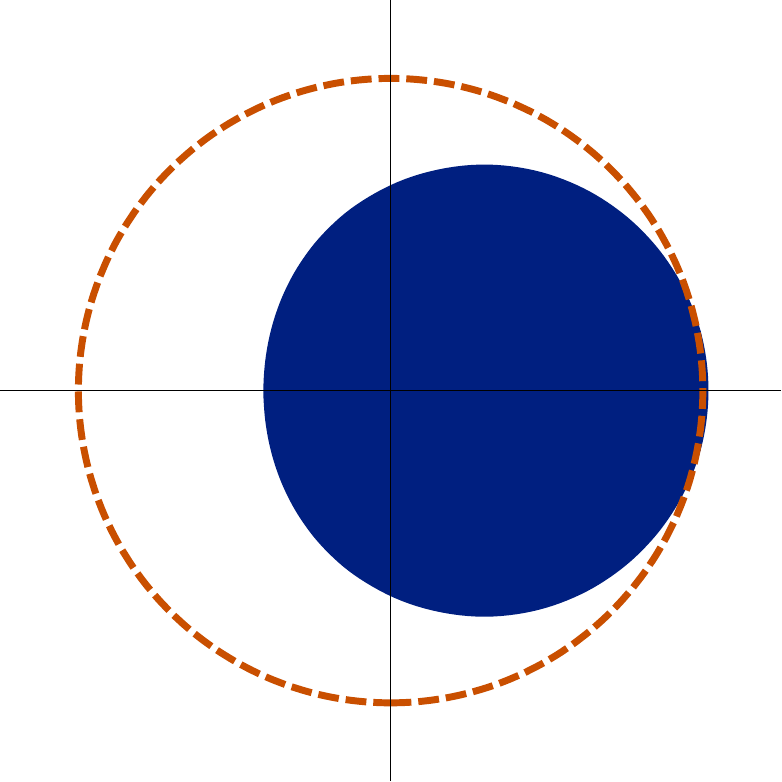}	& 
		\includegraphics[scale=0.4]{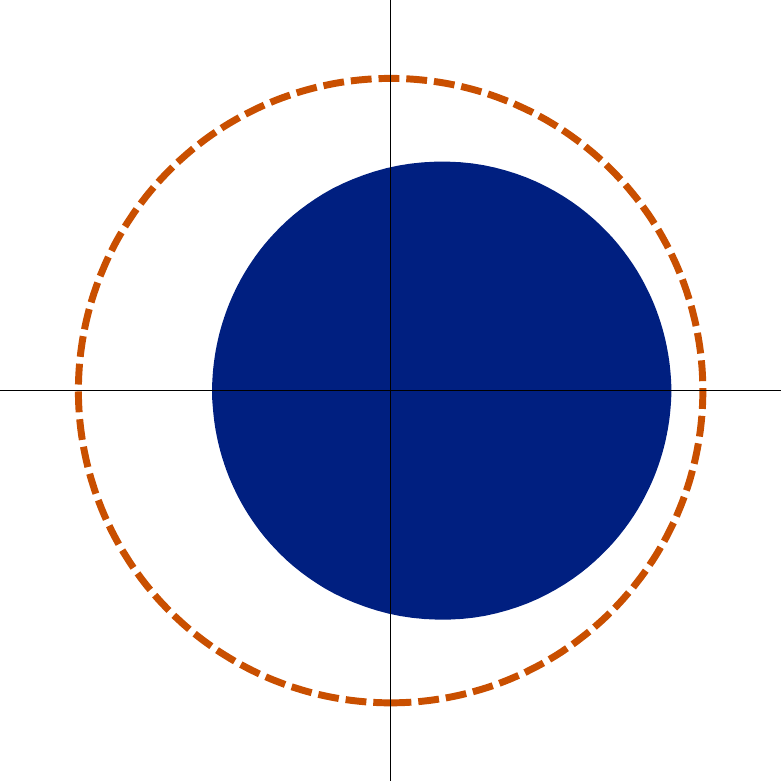}	& 
		\includegraphics[scale=0.4]{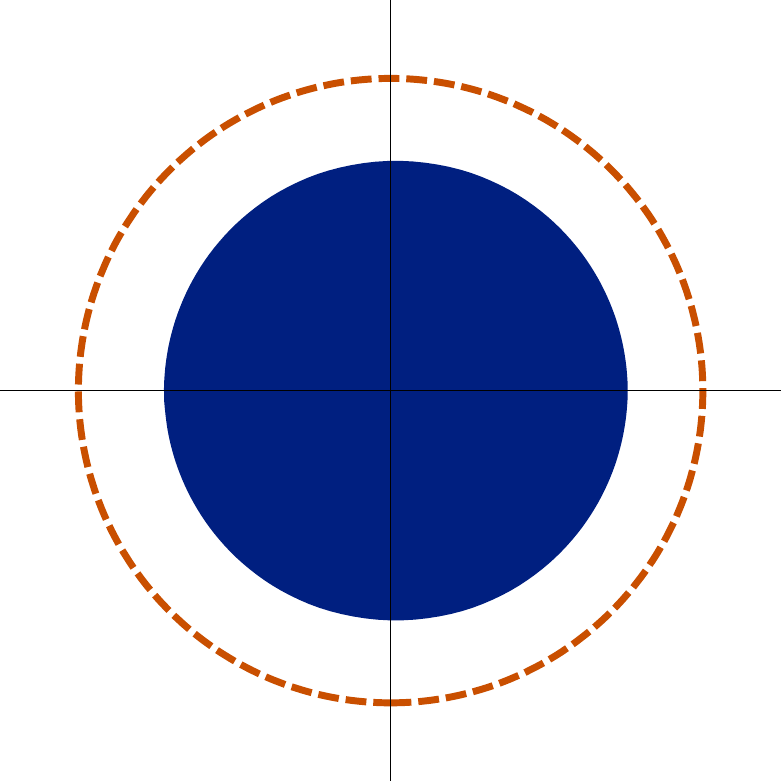}	\\
		$\vartheta_{O}=\frac{\pi}{2}$ & $\vartheta_{O}=\frac{3 \pi}{8}$ & 
		$\vartheta_{O}=\frac{\pi}{4}$ & $\vartheta_{O}=\frac{\pi}{8}$ & 
		$\vartheta_{O}=0$
	\end{tabular}
	\caption[Plots of black hole's shadow for different observer 
		positions]{Shadow of a black hole for an observer at $r_O = 5m$ and
		different inclination angles $\vartheta _O$, with fixed $\beta=\frac{5}{9}m^{2}$, 
		$\ell=\frac{4}{3}m$, $\Lambda = {10}^{-2}m^{-2}$ and 
		$a=a_{\mathrm{max}}\approx 1.51m$. As in Fig.~\ref{fig:Shadow},
    the cross hairs indicate the spatial direction towards the black hole
    and the dashed (red) circle indicates the celestial equator. 
}
	\label{fig:ShadowObserv}
\end{figure*}

In Figs.~\ref{fig:Shadow} and \ref{fig:ShadowObserv} we show 
pictures of the shadow, as it is seen by our chosen observer 
with four-velocity $e_0$. For calculating the boundary curve 
of the shadow we have used our analytical parameter representation, 
and for plotting it we have used stereographic projection from the 
celestial sphere onto a plane, as illustrated in Fig.~\ref{fig:LightObserv}. 
Standard Cartesian coordinates in this plane are given by
\begin{equation}
\begin{aligned}
	x(r_p) &= -2 \tan \Big( \frac{\theta(r_p)}{2} \Big)
		\sin \big( \psi(r_p) \big),\\
	y(r_p) &= -2 \tan \Big( \frac{\theta(r_p)}{2} \Big)
		\cos \big( \psi(r_p) \big).
\end{aligned}
\label{eq:stereo}
\end{equation}

In Fig.~\ref{fig:Shadow} the observer position is kept fixed at 
Boyer--Lindquist coordinates $r_{O}=5 m$ and $\vartheta_{O} = \pi /2$. 
The parameters of the black hole are chosen such that the observer is 
always located in the domain of outer communication. Each of the
five shadings corresponds to a certain choice of parameters
$\beta$, $\ell$ and $\Lambda$, and for each choice 
the shadow is shown for four different values of the spin, 
$a = \lambda a_{\mathrm{max}}$, where $a_{\mathrm{max}}$ is
determined by $\beta$, $\ell$ and $\Lambda$. 
The shadows of the first three cases---Kerr \circc{blau1!20!white},
Kerr--NUT \circc{blau1!40!white}, Kerr--Newman--NUT with cosmological
constant \circc{blau1!60!white}---correspond to the photon regions
presented in Figs.~\ref{fig:Photonregion1}--\ref{fig:Photonregion3}.

We see that the shape of the shadow is largely determined by the 
spin $a$ of the black hole. With increasing $a$ the shadow becomes 
more and more asymmetric with respet to a vertical axis. This 
asymmetry is well-known from the 
Kerr metric and it is easily understood as a ``dragging effect''
of the rotating black hole on the light rays. The other parameters
$\beta$, $\ell$ and $\Lambda$ have an effect on the size of the
shadow but, at least for the naked eye, hardly on its shape. Note
that the size of the shadow depends, of course, on $r_O$ and that
there is no direct way of comparing radius coordinates in 
different space-times operationally. Therefore, if we want to 
get some information on the space-time from observing the shadow,
the shape is much more relevant than the size.

In Fig. \ref{fig:ShadowObserv} we consider an extremal black hole, 
$a=a_{\mathrm{max}}$, with fixed parameters $\beta$, $\ell$ and $\Lambda$.
We keep the radius coordinate $r_O$ of the observer fixed, and we vary 
the inclination $\vartheta _O$. Clearly, the asymmetry with respect
to the vertical axis vanishes if the observer approaches the axis, 
$\vartheta _O \to 0$. We have already emphasized the remarkable fact 
that there is no asymmetry with respect to the horizontal axis.

We should mention that in the case $\ell \neq 0$ some light rays
have to pass through the singularity on the axis. We have assumed
that these light rays are \emph{not} blocked, i.e., that
the source of the gravitomagnetic NUT field does \emph{not}
cast a shadow.

\section{Conclusions and Outlook}

Based on a detailed analysis of the photon regions in black-hole
space-times of the Pleba{\'n}ski class, we have derived an analytical 
formula for the shadows of such black holes. As the space-times
under consideration are not in general asymptotically flat and
may have a cosmological horizon, one cannot restrict to observers
at infinity as it was done in many earlier articles on shadows
of black holes. Our formalism allows for observers at any 
Boyer--Lindquist coordinates in the domain of outer communication.
The boundary curve of the shadow was calculated for observers
with a certain four-velocity $e_0$, given by \eqref{eq:newcoord}.
For these observers, the shadow turned out to be always symmetric
with respect to a horizontal axis, even for non-vanishing NUT
parameter $\ell$ and for an observer off the equatorial plane.
For observers with a four-velocity different from $e_0$, the 
shadow can be easily calculated by combining our results with
the standard aberration formula of special relativity. If this
additional aberration effect is taken into account, the 
boundary curve of the shadow will depend on the parameters 
$a$, $\ell$, $\beta$ and $\Lambda$, on the coordinates
$r_O$ and $\vartheta _O$ of the observer, and on the velocity 
of the observer relative to an observer with four-velocity
$e_0$. (The mass $m$ gives an overall scale, and the Manko-Ruiz
parameter $C$ has no influence on the shadow.) We are planning 
to investigate, in a follow-up article, to what extent all these
parameters can be determined from the boundary curve of the 
shadow. With an analytical formula for the boundary curve at
hand, it is a natural idea to use a Fourier analysis of the 
boundary curve and to see how the parameters of the black hole
can be extracted from the Fourier coefficients.

We have restricted to black-hole space-times, but a large part 
of the material presented in this paper is valid for naked 
singularities as well. In particular, the characterization
of the photon region by inequality \eqref{eq:regionK} is true
in general. A major difference is in the fact that in the case 
of a naked singularity there is no domain of outer communication, 
so the possible observer positions are restricted only by a 
cosmological horizon, if present. The shadow of a naked
singularity is drastically different from the shadow of a 
black hole, as was demonstrated by de Vries \cite{Vries.2000} for the 
Kerr-Newman case. While for a black hole the shadow is two-dimensional
(an area on the sky, bounded by a closed curve), for a 
naked singularity the shadow is one-dimensional (an
arc on the sky).

\section*{Acknowledgments}
We would like to thank Domenico Giulini, Norman G\"urlebeck, Eva
Hackmann, Friedrich Hehl, Valeria Kagramanova, Jutta Kunz, and 
Olaf Lechtenfeld for helpful discussions, and Silke Britzen, Frank 
Eisenhauer, and Heino Falcke for valuable information on the status 
of observations.
We gratefully acknowledge support from the DFG within the Research 
Training Group 1620 \blqq Models of Gravity\brqq\ and from the 
\blqq Centre for Quantum Engineering and Space-Time Research (QUEST)\brqq.

\bibliographystyle{apsrev4-1}
%

%

\end{document}